\documentclass[usenatbib]{mnras}
\usepackage{aas_macros}
\usepackage{natbib}
\usepackage{hyperref}
\usepackage{amsfonts}
\usepackage{amsmath}
\usepackage{amssymb}
\usepackage{graphicx}
\usepackage[dvipsnames]{xcolor}
\hypersetup{colorlinks=true,allcolors=teal}

\newcommand{\p}{\ensuremath{\partial}}

\newcommand{\Msun}{\ensuremath{M_{\odot}}}
\newcommand{\Mh}{\ensuremath{h^{-1}M_{\odot}}}

\newcommand{\Mpch}{\ensuremath{h^{-1}{\rm Mpc}}}
\newcommand{\kpch}{\ensuremath{h^{-1}{\rm kpc}}}

\newcommand{\avg}[1]{\ensuremath{\left\langle \,#1\, \right\rangle}}
\newcommand{\e}[1]{\ensuremath{{\rm e}^{#1}}}

\newcommand{\eqn}[1]{equation~\eqref{#1}}

\newcommand{\be}{\begin{equation}}
\newcommand{\ee}{\end{equation}}
\newcommand{\Cal}[1]{\ensuremath{\mathcal{#1}}}

\title[Cosmic web and halo assembly bias]
      {Cosmic web anisotropy is the primary indicator of halo assembly bias} 

\date{draft}

\author[Ramakrishnan et al.]{
Sujatha Ramakrishnan$^{1}$\thanks{E-mail: rsujatha@iucaa.in}, 
Aseem Paranjape$^{1}$\thanks{E-mail: aseem@iucaa.in}, 
Oliver Hahn$^{2}$\thanks{E-mail: oliver.hahn@oca.eu} 
\& Ravi K. Sheth$^{3,4}$\thanks{E-mail: shethrk@physics.upenn.edu}
\\  
 $^1$ Inter-University Centre for Astronomy \& Astrophysics,
      Ganeshkhind, Post Bag 4, Pune 411007, India\\
 $^2$ Laboratoire Lagrange, Universit\'e C\^ote d'Azur, Observatoire de la C\^ote d'Azur, CNRS, Blvd de l'Observatoire,\\\hskip0.15in CS 34229, 06304 Nice cedex 4, France\\
 $^3$ Center for Particle Cosmology, University of Pennsylvania, 209 S. 33rd St., Philadelphia, PA 19104, USA\\
 $^4$ The Abdus Salam International Center for Theoretical Physics, Strada Costiera, 11, Trieste 34151, Italy}

\begin{document}

\label{firstpage}
\pagerange{\pageref{firstpage}--\pageref{lastpage}}

\maketitle

\begin{abstract}
\noindent
The internal properties of dark matter haloes correlate with the large-scale halo clustering strength at fixed halo mass -- an effect known as assembly bias -- and are also strongly affected by the local, non-linear cosmic web. 
Characterising a halo's local web environment by its tidal anisotropy $\alpha$ at scales $\sim4\,\times\,$the halo radius, we demonstrate that these multi-scale correlations represent \emph{two distinct statistical links}: one between the internal property and $\alpha$, and the other between $\alpha$ and large-scale ( $\gtrsim30\Mpch$) halo bias $b_1$. 
We focus on scalar internal properties of haloes related to formation time (concentration $c_{\rm vir}$), shape (mass ellipsoid asphericity $c/a$), velocity dispersion structure (velocity ellipsoid asphericity $c_v/a_v$ and velocity anisotropy $\beta$) and angular momentum (dimensionless spin $\lambda$) in the mass range $8\times10^{11}\lesssim M_{\rm vir}/(\Mh)\lesssim5\times10^{14}$. 
Using conditional correlation coefficients and other detailed tests, we show that the joint distribution of $\alpha$, $b_1$ and \emph{any} of the internal properties $c\in\{\beta,c_v/a_v,c/a,c_{\rm vir},\lambda\}$ is consistent with $p(\alpha,b_1,c)\simeq p(\alpha)p(b_1|\alpha)p(c|\alpha)$, at all but the largest masses. 
\emph{Thus, the assembly bias trends $c\leftrightarrow b_1$ reflect the two fundamental correlations $c\leftrightarrow\alpha$ and $b_1\leftrightarrow\alpha$.}  
Our results are unaffected by the exclusion of haloes with recent major merger events or splashback objects, although the latter are distinguished by the fact that $\alpha$ \emph{does not} explain their assembly bias trends. 
The overarching importance of $\alpha$ provides a new perspective on the nature of assembly bias of distinct haloes, with potential ramifications for incorporating realistic assembly bias effects into mock catalogs of future large-scale structure surveys and for detecting galaxy assembly bias. 
\end{abstract}

\begin{keywords}
cosmology: theory, dark matter, large-scale structure of the Universe -- methods: numerical
\end{keywords}

\section{Introduction}
\label{sec:intro}
\noindent
The physical connection between the growth and properties of gravitationally collapsed dark matter haloes and the cosmic web environment in which these haloes reside is an interesting and challenging problem in the study of hierarchical structure formation \citep{ws79,el95,bm96,bkp96,monaco99,st99}. Although the basic statistical connection between the very large-scale density environment (or halo bias) and halo properties such as mass was already established several decades ago \citep{kaiser84,bbks86,bcek91,lc93}, subsequent technological improvements in simulating cold, collisionless self-gravitating cosmological systems have revealed several additional features of dark matter haloes.

Primarily, these relate to the striking universality seen in the structure of cold dark matter (CDM) haloes, both in the density \citep{nfw96,nfw97} as well as velocity dispersion profiles \citep{tn01,ludlow+10}. Later results also indicate a deep connection -- which is the focus of this work -- between the large-scale halo bias and internal properties of haloes of fixed mass such as formation time, concentration, substructure abundance, shape, velocity dispersion structure, angular momentum, etc. \citep[see, e.g.,][]{st04,gsw05,wechsler+06,jsm07,fw10}. Apart from the intrinsic interest in painting a more complete picture of hierarchical structure formation from first principles, understanding and calibrating these effects also continues to be of interest from the point of view of galaxy formation and evolution \citep[see, e.g.,][]{yfw13,lin+16,tinker+17,phs18b,azpm18,wang+18,zehavi+18} as well as for precision cosmology \citep{zhv14,mw18}. 

The dependence of halo bias on halo formation time at fixed mass was termed `assembly bias' in the early literature on this subject. We will use this term to denote the dependence of bias on any internal property other than mass, although recent results indicate that there could be more than one physical mechanism responsible for establishing these correlations \citep[see, e.g.,][]{mzw18,salcedo+18,han+19}.

In general, such correlations between internal halo properties (i.e., quantities defined at length scales $\lesssim\textrm{few}\times100\kpch$, say) and large-scale halo bias (measured at scales $\gtrsim\textrm{few}\times10\Mpch$) can be thought of as remnants of the physics of halo formation in the hierarchical paradigm. For example, excursion set models of halo abundances and clustering generically predict such statistical correlations by connecting the local physics of halo formation to the large-scale halo environment through the long-wavelength correlations present in the initial conditions \citep[see, e.g.,][]{zentner07,dwbs08,ms12}. These models, however, currently do not correctly reproduce all known assembly bias trends, indicating that they still lack some key physical mechanisms involved in halo formation. 

Focusing on the expected local and highly non-linear nature of halo formation, it is then interesting to ask whether one might segregate the correlation between an internal property and large-scale bias into (at least) two distinct contributions: one composed of a connection between the internal property and some feature of the \emph{local} non-linear environment and the other connecting the local environment to the large-scale bias. The latter connection is conceptually exactly the kind of correlation that excursion set models are built to explain, while the former could be a correlation which needs additional physical mechanisms to be included in the dynamical models describing halo formation.

Recent studies indicate that the \emph{cosmic web environment} at relatively small scales (of the order of a few virial radii) plays an important role in the assembly bias due to halo formation epoch \citep{hahn+09}, mass accretion rate \citep{fm10,musso+18}, internal velocity dispersion structure \citep{bprg17} and halo concentration \citep{phs18a}. 
In particular, these studies have revealed an intimate connection between the nature of assembly bias and the immediate environment of a halo \citep[e.g., whether or not the halo lives in a cosmic filament; see also][who studied the dependence of dynamical variables on the local tidal environment]{wang+11,swm15}. While this is not unexpected -- the protohalo patches from which haloes form are correlated with the linear tidal field \citep{bm96,smt01} -- and analytical excursion set calculations do predict a statistical correlation between halo bias and formation time or concentration \citep{ms12,cs13}, the specific role of the non-linear cosmic web in establishing assembly bias effects still lacks a first principles understanding.

In this work, we are interested in clean statistical signatures, using $N$-body simulations, that the physics of hierarchical halo formation splits into distinct contributions from different length scales.
Previous work by some of us has shown that the \emph{tidal anisotropy} in the immediate vicinity of a halo (see below) plays a key role in determining the assembly bias trends defined by halo concentration, particularly at low masses where a large fraction of haloes reside in filaments \citep{phs18a}. 
This appears quite natural in hindsight, since the turn-around radius of material currently infalling onto a halo is a few times the halo radius \citep[around the same scale where][defined the tidal anisotropy]{phs18a}, and the only relevant physical mechanism at play for collisionless dark matter is the tidal influence of gravity.
Our goal here is to extend these ideas to other halo properties (we will study the halo shape, velocity dispersion tensor and spin) and \emph{statistically assess the importance of the tidal anisotropy as an intermediary in explaining assembly bias in these properties.}

The paper is organised as follows. In section~\ref{sec:sims}, we describe our simulations and the measurements of various internal properties of haloes used in this work. In section~\ref{sec:correlns}, we explore the connection between the halo tidal environment and assembly bias in these properties. We summarize known results before presenting our main findings which indicate that the tidal anisotropy of the cosmic web in the halo vicinity is an important indicator of \emph{all} assembly bias trends. In section~\ref{sec:splashbackmergers}, we present tests of potential physical explanations of our results, showing that the connection between tidal anisotropy and assembly bias \emph{cannot be explained} by splashback objects or recent mergers. We conclude with a discussion in section~\ref{sec:conclude}. The Appendix presents convergence studies for our numerical techniques and detailed tests of the robustness of our choice of statistics.

Throughout, we use a spatially flat $\Lambda$CDM cosmology with total matter density parameter $\Omega_{\rm m}=0.276$, baryonic matter density $\Omega_{\rm b}=0.045$, Hubble constant $H_0=100h\,{\rm kms}^{-1}{\rm Mpc}^{-1}$ with $h=0.7$, primordial scalar spectral index $n_{\rm s}=0.961$ and r.m.s. linear fluctuations in spheres of radius $8\Mpch$, $\sigma_8=0.811$, with a transfer function generated by the code \textsc{camb} \citep{camb}.\footnote{\href{http://camb.info}{http://camb.info}}

\section{Simulations and halo properties}
\label{sec:sims}
\noindent
We use $N$-body simulations of collisionless CDM in cubic, periodic boxes performed using the tree-PM code \textsc{gadget-2} \citep{springel:2005}.\footnote{\href{http://www.mpa-garching.mpg.de/gadget/}{{http://www.mpa-garching.mpg.de/gadget/}}} These simulations, which we briefly describe here, are the same as those used by \citet{phs18a} in their analysis. We use two configurations: a lower resolution one having 10 independent realisations, and 2 realisations of a smaller volume, higher resolution box. All boxes were run using $N_{\rm p} = 1024^3$ particles, with the lower (higher) resolution configuration having a box of comoving length $L = 300 \,(150)\Mpch$, corresponding to a particle mass of $m_{\rm p} = 1.93 \times 10^9 \,(2.4 \times 10^8) h^{-1}\Msun$. The force resolution parameter $\epsilon$ in each case was set to $1/30$ of the mean comoving inter-particle spacing, leading to $\epsilon=9.8\,(4.9) \kpch$ for the lower (higher) resolution, while PM forces were computed on a $2048^3$ grid in each case.

Initial conditions for the lower (higher) resolution boxes were generated at a starting redshift $z_{\rm in}=49\,(99)$ using the code \textsc{music} \citep{hahn11-music}\footnote{\href{https://www-n.oca.eu/ohahn/MUSIC/}{https://www-n.oca.eu/ohahn/MUSIC/}} with 2nd order Lagrangian perturbation theory. Haloes were identified using the code \textsc{rockstar} \citep{behroozi13-rockstar}\footnote{\href{https://bitbucket.org/gfcstanford/rockstar}{https://bitbucket.org/gfcstanford/rockstar}} which performs a Friends-of-Friends (FoF) algorithm in 6-dimensional phase space. For the higher resolution boxes, we stored $201$ snapshots equally spaced in the scale factor $a=1/(1+z)$ ($\Delta a=0.004615$) between $z=12$ and $z=0$, which we used to produce merger trees using the code \textsc{consistent-trees} \citep{behroozi13-consistenttrees}.\footnote{\href{https://bitbucket.org/pbehroozi/consistent-trees}{https://bitbucket.org/pbehroozi/consistent-trees}}
The simulations and analysis were performed on the Perseus cluster at IUCAA.\footnote{\href{http://hpc.iucaa.in}{http://hpc.iucaa.in}}

To ensure that our results are not contaminated by substructure and numerical artefacts, we discard all subhaloes identified by \textsc{rockstar} and further only consider objects whose virial energy ratio $\eta = 2T /|U|$ satisfies $0.5 \leq \eta \leq 1.5$ as suggested by \citet{bett+07}. Our measurements of the tidal environment in the vicinity of the haloes, which we describe below, were performed after Gaussian smoothing on a cubic grid with $N_{\rm g} = 512^3$ cells. We therefore impose a restriction on the minimum halo mass we study, so as to minimise the contamination to our final results from the resolution imposed by this grid. Based on convergence studies which we discuss below, we choose to analyse haloes with at least $3200$ particles for most of the analysis. This gives a mass threshold of $6.2 \times 10^{12} \,(7.7\times 10^{11}) \Mh$ and leaves us with approximately $19,000$ $(17,000)$ objects in each of the lower (higher) resolution boxes. 

Throughout, we focus on results at $z = 0$ and study all halo properties as a function of virial mass $M_{\rm vir}$ enclosed in the virial radius $R_{\rm vir}$ as defined using the spherical overdensity prescription of \citet{bn98}. 
We have checked that qualitatively identical results are obtained when binning haloes according to other mass definitions such as $M_{\rm 200b}$ enclosed inside the radius $R_{\rm 200b}$,\footnote{$R_{\rm 200b}$ is the halo-centric radius which encloses a spherical overdensity of $200$ times the background matter density.} or the mass $M_{\rm ell}$ enclosed inside the mass ellipsoid of the halo which is calculated as described in section~\ref{subsubsec:massellipsoid} below.

\subsection{Measuring halo-by-halo bias}
\label{subsec:b1}
\noindent
As our indicator of choice for the large-scale density environment of haloes, we use the halo-by-halo bias estimator $b_1$ described by \citet{phs18a}. A similar variable defined in real space has also been recently used by \citet{han+19}.

This is essentially a halo-centric dark matter overdensity estimate filtered with a window function that is sharp in Fourier space. This sharp-$k$ filter is built using $k$-dependent weights chosen such that the arithmetic mean of $b_1$ for any population of haloes is identical to the usual Fourier space linear bias of this population, as measured by the ratio of the halo-matter cross power spectrum $P_{\rm hm}(k)$ and the matter power spectrum $P_{\rm mm}(k)$ at small $k$.

In detail, denoting the discrete Fourier transform of the dark matter density contrast as $\delta(\mathbf{k})$ evaluated at the grid location $\mathbf{k}$ in Fourier space, the bias for halo $h$ is calculated as
\begin{align}
b_{1,h} &= \sum_{\textrm{low-}k} \,w_k\bigg[ \avg{\e{i\mathbf{k}\cdot\mathbf{x}(h)}\delta^\ast(\mathbf{k})}_k/P_{\rm mm}(k) \bigg] \,,
\label{eq:b1HbyH}
\end{align}
where $P_{\rm mm}(k) = \avg{\delta(\mathbf{k})\delta^\ast(\mathbf{k})}_k$ and $\avg{\ldots}_k$ denotes a spherical average over modes contained in a bin of $k$. The quantity \e{i\mathbf{k}\cdot\mathbf{x}(h)} corresponds to a weighted average of phase factors over the configuration space cell $\mathbf{x}(h)$ containing the halo $h$, and $7$ of its neighbouring cells, using weights appropriate for a cloud-in-cell (CIC) interpolation. We sum over low-$k$ modes in the simulation box, using the ranges  $0.025 (0.05)\lesssim k/(h{\rm Mpc}^{-1})\lesssim0.09$ for the lower (higher) resolution configuration, additionally weighting by the number of modes $w_k\propto k^3$ for logarithmically spaced bins (with $\sum_{\textrm{low-}k}w_k=1$).

\emph{We emphasize that the resulting bias estimate is an indicator of halo environment at large scales $\gtrsim30\Mpch$ where bias is approximately linear and scale-independent.} This should be contrasted with other estimators employed in the literature, such as marked correlation functions or ratios of correlation functions at scales $\lesssim10\Mpch$ \citep{wechsler+06,villarreal+17,mk19}. The interpretation of assembly bias trends of these estimators is likely to be complicated by non-linearity and/or scale-dependence of bias \citep{shpl16,pp17}. See also \citet{han+19} for tests of linearity at smaller scales.

The primary advantage of using a halo-by-halo estimator of bias is that it allows us to treat halo bias on par with any other halo-centric or internal property. In particular, we are able to directly probe the correlation of the scatter in halo bias with other variables by calculating appropriate correlation coefficients between $b_1$ and these variables, \emph{without} having to bin haloes. We will build our main analysis below using such correlation coefficients.

\subsection{Measuring the halo tidal environment}
\label{subsec:tidalmsrmnt}
As our main indicator of a halo's non-linear local environment, we will use the \emph{tidal anisotropy} variable $\alpha$ introduced by \citet{phs18a}. This is constructed using measurements of the tidal tensor at halo locations, as follows.

First, the density field $\delta(\mathbf{x})$ evaluated using CIC interpolation on a cubic lattice is used to evaluate the tidal tensor $\psi_{ij}(\mathbf{x})\equiv\p^2\psi/\p x^i\p x^j$ by inverting the normalised Poisson equation $\nabla^2\psi=\delta$ in Fourier space. While doing so, we apply a range of Gaussian smoothing filters $\e{-k^2R_{\rm G}^2/2}$ to generate multiple smoothed versions $\psi_{ij}(\mathbf{x};R_{\rm G})$ of the tidal tensor on the lattice. We then interpolate these in configuration space to the location $\mathbf{x}_h$ of halo $h$ and also interpolate in smoothing scales to the size $R_h$ of the halo (see below), thus creating a halo-by-halo catalog of tidal tensor estimates $\psi_{ij}(\mathbf{x}_h;R_h)$.

Diagonalising this halo-centric tidal tensor and denoting its eigenvalues by $\lambda_1\leq\lambda_2\leq\lambda_3$ (for brevity, we will drop the subscript $h$ in the following), we then construct the halo-centric overdensity $\delta$ using
\be
\delta = \lambda_1 + \lambda_2 + \lambda_3\,,
\label{eq:delta-def}
\ee
and the halo-centric tidal shear $q^2$ using \citep{hp88,ct96a}
\be
q^2 = \frac12\left[\left(\lambda_2-\lambda_1\right)^2 + \left(\lambda_3-\lambda_1\right)^2 + \left(\lambda_3-\lambda_2\right)^2\right]\,.
\label{eq:q2-def}
\ee
The halo-centric tidal anisotropy $\alpha$ is then defined by
\be
\alpha = \sqrt{q^2}/\left(1+\delta\right)\,.
\label{eq:alpha-def}
\ee
The choice of smoothing scale $R_{\rm G}=R_h$ for each halo is driven by our requirement of a measure of the local halo tidal environment which correlates well with the \emph{large-scale} environment as measured by $b_1$ above. As shown by \citet{phs18a}, the choice $R_h \sim 4R_{\rm 200b}$ is the largest halo-scaled smoothing radius\footnote{In practice, we set $R_h=4R_{\rm 200b}/\sqrt{5}$, the ``Gaussian equivalent'' of the spherical tophat scale $4R_{\rm 200b}$. The factor $\sqrt{5}$ is most easily understood by Taylor expanding the Fourier transforms of the Gaussian and spherical tophat filters and equating the terms proportional to $k^2$.} for which $\alpha$ as defined above correlates more tightly with $b_1$ than does $\delta$ at the same scale (see also Appendix~\ref{app:tidal-bias}).

The measurements of the tidal tensor and associated variables above depend on the choice of grid size used for the original CIC interpolation. For a given grid size, the requirement that the sphere of radius $\sim4R_{\rm 200b}$ be sufficiently well-resolved leads to a lower limit on halo mass. Appendix~\ref{app:convergence} presents a convergence study using which we conclude that a $512^3$ grid is sufficient for our purposes, provided we restrict attention to haloes with $\geq3200$ particles enclosed inside $R_{\rm vir}$. These are the default choices for our analysis.

Figure~\ref{fig:convergence} also shows that $\alpha$ and $\delta$ as defined above are, in fact, positively correlated. This is potentially a cause for concern because any statements regarding the correlation between $\alpha$ and halo properties could simply be reflecting a correlation between $\delta$ and those properties \citep[see, e.g.,][]{ss18}. To assess the level to which this is true, we perform a detailed comparison of these correlations in Appendices~\ref{app:tidal-bias} and~\ref{app:tidal-internal}, finding that $\alpha$ is in fact a better indicator of \emph{all} correlations with halo properties than is $\delta$. (We remind the reader that we define both $\alpha$ and $\delta$ at scales $\sim4R_{\rm 200b}$ for the reasons discussed above.)

We also note that other estimators of tidal anisotropy such as $\sqrt{q^2}/(1+\delta)^\mu$ with some constant $\mu$ can decrease the correlation strength between $\alpha$ and $\delta$. E.g., \citet{azpm18} find that setting $\mu\simeq0.55$ works well for $R_{\rm G}=5\Mpch$ and haloes selected so as to describe a sample of galaxies in the Sloan Digital Sky Survey. However, the dependence of the value of $\mu$ on smoothing scale, halo mass, large-scale environment or sample selection, and the origin of any specific value, is unclear. We therefore prefer to work with our definition \eqref{eq:alpha-def}, which is a regular function of $1+\delta$, and explicitly check for systematic biases due to correlations with $\delta$. 

For example, since $\alpha$ and $\delta$ are positively correlated, one might ask whether the variable $\alpha_{(2)}\equiv\sqrt{q^2}/(1+\delta)^2$, which is also a regular function of $1+\delta$, might perform better. Indeed, we find that $\alpha_{(2)}$ correlates very weakly with $\delta$ over our entire mass range \citep[see also][for an alternative tidal variable which also correlates weakly with the isotropic overdensity]{haas+12}. However, the $b_1\leftrightarrow\alpha_{(2)}$ correlation is weaker than the $b_1\leftrightarrow\alpha$ correlation, and is instead similar to the $b_1\leftrightarrow\delta$ correlation seen in Figure~\ref{fig:tidal-bias}, thus making $\alpha_{(2)}$ unsuitable for our purposes. Thus, although the tidal anisotropy variable $\alpha$ as defined in \eqn{eq:alpha-def} is strictly a combination of anisotropy and density, its superior correlation with $b_1$ as compared to pure anisotropy (or pure density) variables makes $\alpha$ our variable of choice for assembly bias studies.

\subsection{Measuring internal halo properties}
\noindent
We will study the correlations between the halo environment (as characterised by halo bias $b_1$ and tidal anisotropy $\alpha$) and a number of internal halo properties. For the latter, we will focus on \emph{scalar} variables describing the anisotropy of the halo shape and velocity dispersion tensors, halo concentration and spin. We discuss the measurements of each of these below. 

Throughout this work, for any halo \emph{we discard particles that are either not contained inside the phase space FoF grouping provided by \textsc{rockstar} or are gravitationally unbound to the halo.} All internal halo properties are therefore calculated using only gravitationally bound particles belonging to the FoF group of each halo.

\subsubsection{Mass ellipsoid tensor}
\label{subsubsec:massellipsoid}
\noindent
As a part of its post-processing analysis, the \textsc{rockstar} code measures the mass ellipsoid tensor (or shape tensor) $M_{ij}$ of each halo using the iterative procedure prescribed by \citet{allgood+06}. This tensor is evaluated as
\be
M_{ij} = \sum_{n\in\textrm{halo}}\,x_{n,i}x_{n,j} / r_n^2
\label{eq:massellipsoidtensor}
\ee
where $i,j=1,2,3$ refer to the coordinate directions, $\mathbf{x}_n$ is the comoving position of the $n^{\rm th}$ particle in the halo \emph{with respect to the halo center of mass} and $r_n^2$ is the comoving ellipsoidal distance of this particle from the center of mass given by $r_n^2 = x_n^2 + y_n^2/(b/a)^2 + z_n^2/(c/a)^2$. Here, we defined $a^2\geq b^2\geq c^2$ as the ordered eigenvalues of $M_{ij}$. Since the calculation of the ellipsoidal distance requires knowledge of the eigenvalue ratios, this is done by an iterative procedure with a starting guess of equal eigenvalues and subsequent updates in each iteration after estimating $M_{ij}$ using \eqn{eq:massellipsoidtensor} and diagonalising it. The calculation sets the semi-major axis of the ellipsoid equal to the halo virial radius $R_{\rm vir}$ and sums over all (bound, FoF) particles in the halo. We refer the reader to \citet{allgood+06} for further details of the procedure.

Denoting the final converged eigenvalues with the same notation $a^2\geq b^2\geq c^2$, we use the ratio $c/a$ as a measure of the asphericity of the mass ellipsoid tensor. This variable is convenient since its values are bounded between $0\leq c/a \leq 1$, with zero corresponding to a highly aspherical halo and unity to a spherical halo. We have checked that using other measures of asphericity which include information on the intermediate axis, such as the triaxiality variable $\Cal{T} = (a^2-b^2)/(a^2-c^2)$ \citep{fiz91}, lead to qualitatively identical results.

\subsubsection{Velocity ellipsoid tensor}
\label{subsubsec:velocityellipsoid}
\noindent
We have modified \textsc{rockstar} so as to calculate the velocity ellipsoid tensor which is a measure of the anisotropic velocity dispersion of the dark matter particles constituting a halo. For a halo with $N$ particles, this tensor is given by
\be
V^2_{ij} = \frac1N\sum_{n\in\textrm{halo}} \left(v_{n,i}-\avg{v_i}\right)\left(v_{n,j}-\avg{v_j}\right)
\label{eq:velocityellipsoidtensor}
\ee
where $\mathbf{v}_{n}$ is the peculiar velocity of $n^{\rm th}$ dark matter particle and $\avg{\mathbf{v}} = \sum_{n\in\textrm{halo}}\mathbf{v}_n/N$ is the bulk peculiar velocity of the halo. 

Similarly to the mass ellipsoid tensor, we denote the eigenvalues of $V^2_{ij}$ by $a_v^2\geq b_v^2\geq c_v^2$ and use the ratio $c_{v}/a_{v}$ as a measure of the asphericity of the velocity ellipsoid. For consistency with the calculation of the mass ellipsoid tensor, we restrict the sum in \eqn{eq:velocityellipsoidtensor} to be over those (bound, FoF) particles contained inside the mass ellipsoid defined by \eqn{eq:massellipsoidtensor}. 

\subsubsection{Velocity anisotropy}
\label{subsubsec:beta}
\noindent
A related property of the halo is the velocity anisotropy $\beta$ defined as \citep[e.g., ][]{binney-tremaine-GalDyn}
\be
\beta = 1 -\sigma_{\rm t}^2/\left(2\sigma_{\rm r}^2\right)\,,
\label{eq:velocityanisotropy}
\ee
where $\sigma_{\rm r}^2$ and $\sigma_{\rm t}^2$ are the radial and tangential velocity dispersion, respectively, of the particles in the halo. These are calculated by first projecting the velocity of each particle in the halo along and perpendicular to the radial direction (defined by the center of mass) and then computing the variance of each component separately over all particles. As before, we restrict attention to the particles contained inside the mass ellipsoid defined by \eqn{eq:massellipsoidtensor}.
We have modified \textsc{rockstar} to compute $\beta$ for each halo alongside the velocity and mass ellipsoid calculations described previously.

Although $\beta$ is clearly related to the velocity ellipsoid tensor, it is worth keeping in mind that $\beta$ also crucially depends on the \emph{shape} of the halo when computing the radial and tangential dispersions. Thus, the velocity anisotropy $\beta$ captures information from the full phase space of the halo, unlike the mass and velocity ellipsoid tensors individually. We return to this point below. For now, we note that this variable takes values in the range $-\infty<\beta\leq1$, with $\beta=0$ corresponding to an isotropic velocity ellipsoid and the positive and negative extremes of the allowed range corresponding, respectively, to radially and tangentially dominated velocity dispersions.

Finally, unlike standard applications which study $\beta$ as a function of radial distance, here we define the radial and tangential dispersions, and hence $\beta$, by averaging over all (bound, FoF) particles in the halo. It would also be interesting to explore the radial dependence of $\beta$ \emph{vis a vis} the environmental correlations we are focusing on, an exercise we leave for future work.

\subsubsection{Concentration}
\label{subsubsec:cvir}
\noindent
By default, \textsc{rockstar} performs a least-squares fit of the spherically averaged dark matter profile of each halo to the universal NFW form \citep*{nfw97}
\be
\rho(r) = \frac{\rho_{\rm s}}{(r/r_{\rm s})\left(1+r/r_{\rm s}\right)^2}\,,
\label{eq:nfwprofile}
\ee
where $\rho_{\rm s}$ is a normalisation constant related to the mass of the halo and $r_{\rm s}$ is the scale radius. The halo concentration $c_{\rm vir}$ is then defined as
\be
c_{\rm vir} \equiv R_{\rm vir} / r_{\rm s} \,.
\label{eq:cvir-def}
\ee
Halo concentration correlates well with formation epoch \citep{nfw97,wechsler+02,ludlow+13}, and its dependence on halo mass and environment has been thoroughly studied in the literature \citep[][see also below]{bullock+01,ludlow+14,dk15}. We include $c_{\rm vir}$ in our analysis as a proxy for formation epoch and to compare with the assembly bias trends of other variables.

\subsubsection{Spin}
\label{subsubsec:spin}
\noindent
The dimensionless spin parameter is given by
\begin{equation}
\lambda \equiv \frac{J |E|^{1/2}}{ G M_{\rm vir}^{5/2}}
\label{eq:lambda-def}
\end{equation}
where $J$ is the magnitude of the angular momentum, $E$ the total energy and $M_{\rm vir}$ the mass of the halo, with $G$ being Newton's constant \citep{peebles69}. By default, \textsc{rockstar} calculates $\lambda$ for each halo using its bound, FoF particles inside $R_{\rm vir}$; we use this measurement in our analysis below. 

We have also checked that using the alternative definition of dimensionless spin $\lambda^\prime$ proposed by \citet[][this is also calculated by \textsc{rockstar}]{bullock+01b} leads to identical results, where
\begin{equation}
\lambda' \equiv \frac{J_{\rm vir}}{ \sqrt{2} M_{\rm vir}R_{\rm vir}V_{\rm vir}}
\label{eq:lambdaBullock-def}
\end{equation}  
with $J_{\rm vir}$ being the angular momentum inside a sphere of radius $R_{\rm vir}$ containing mass $M_{\rm vir}$, and where $V_{\rm vir}=\sqrt{GM_{\rm vir}/R_{\rm vir}}$ is the halo circular velocity at radius $R_{\rm vir}$.

\begin{figure*}
\centering
\includegraphics[width=0.85\textwidth]{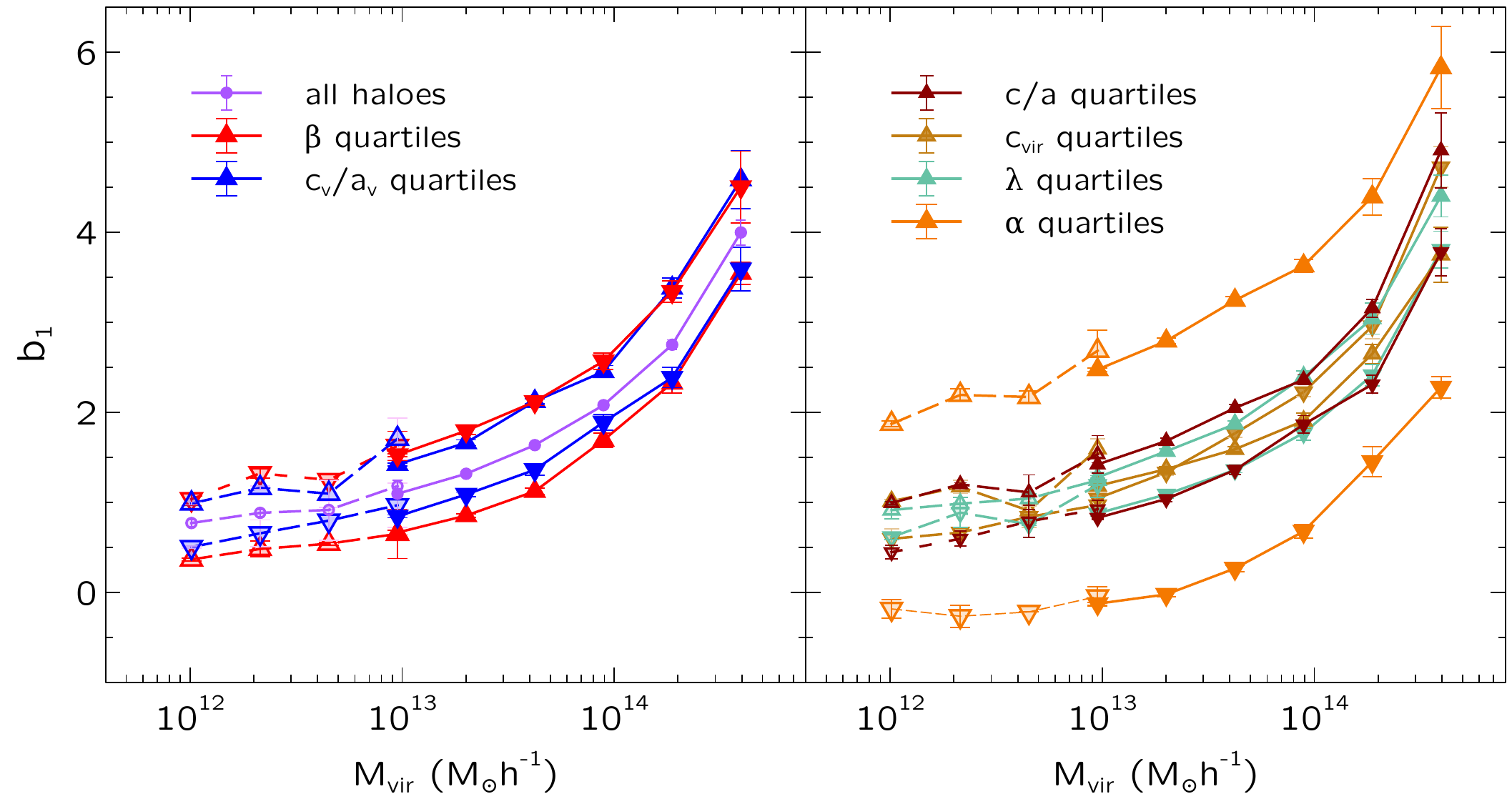}
\caption{
{\bf Summary of known assembly (or secondary) bias trends.}
Symbols joined by lines show measurements of halo bias $b_1$ (section~\ref{subsec:b1}) averaged over haloes in bins of mass $M_{\rm vir}$ for different populations. 
Circles in the \emph{left panel} show results for the full halo population in each mass bin.
Triangles of different colours in each panel indicate measurements at fixed mass but focusing on haloes in the upper quartile (upward triangles) and lower quartile (downward triangles) of a secondary property. 
The \emph{left panel} shows results for the secondary property being velocity anisotropy $\beta$ (section~\ref{subsubsec:beta}) and velocity ellipsoid asphericity $c_v/a_v$ (section~\ref{subsubsec:velocityellipsoid}).  
The \emph{right panel} shows results for halo shape asphericity $c/a$ (section~\ref{subsubsec:massellipsoid}), concentration  $c_{\rm vir}$ (section~\ref{subsubsec:cvir}), spin $\lambda$ (section~\ref{subsubsec:spin}) and the tidal anisotropy $\alpha$ (section~\ref{subsec:tidalmsrmnt}).
In each panel, filled symbols joined with solid lines show the mean over 10 realisations of the lower resolution box, with error bars showing the scatter around the mean, while open symbols joined with dashed lines show measurements using 2 realisations of the higher resolution box. We see that the tidal anisotropy $\alpha$ has, by far, the strongest trend with halo bias at fixed mass.}
\label{fig:assemblybias}
\end{figure*}

Similarly to halo concentration, the distribution of spin as a function of halo mass and its correlation with other halo properties as well as large-scale environment is also well-studied in the literature \citep[e.g.,][see also below]{bullock+01b,bett+07,rodriguez-puebla+16,johnson+19}. The measurement of the spin parameter is rather sensitive to the particle resolution, with order unity errors accrued for haloes sampled with a few hundred particles \citep{onorbe+14,benson17}, and this can in principle substantially affect any conclusions regarding correlations between spin and other variables. Since we only consider haloes sampled with $\geq3200$ particles, however, we expect these numerical errors in our bins of lowest particle count to be $\lesssim25\%$ at the object-by-object level \citep[see Figure~3 of][]{benson17}. We therefore do not expect any of our conclusions regarding spin assembly bias to be altered as a consequence of particle resolution.

Figure~\ref{fig:histogram} shows the distributions of each of these variables for a few narrow mass ranges. See Appendix~\ref{app:internal} for a discussion of the associated trends.

\section{Assembly bias and tidal environment}
\label{sec:correlns}
\noindent
In this section, we use measurements of halo bias, tidal anisotropy and the various internal halo properties discussed in the previous section to assess the nature of the statistical correlations between all these quantities. We start by using our simulations to recapitulate some known results on assembly bias, followed by our new statistical analysis.

\subsection{Known results}
\label{subsec:knownresults}
\noindent
Figure~\ref{fig:assemblybias} summarizes previously known assembly bias / secondary bias trends due to halo velocity anisotropy variables $\beta$ and $c_v/a_v$ \emph{(left panel)} and halo shape $c/a$, concentration $c_{\rm vir}$, spin $\lambda$ and tidal anisotropy $\alpha$ \emph{(right panel)}. In each panel, upward (downward) triangles indicate the mean halo bias in the upper (lower) quartiles of the respective quantity, at fixed halo mass. Additionally, the circles in the left panel show the mean bias for all haloes at fixed mass.

We see that haloes that are aspherical either in shape (small $c/a$) or velocity dispersion (small $c_v/a_v$) are less clustered than more spherical haloes. The split by velocity anisotropy $\beta$ shows that haloes dominated by more radial orbits ($\beta>0$) are less clustered than tangentially dominated haloes. Correspondingly, haloes with smaller spin values are less clustered than those with higher spin. The split by halo concentration shows a more complex trend, with highly concentrated haloes being less clustered at high masses but more clustered at low masses, the inversion occurring near $M_{\rm vir}\sim10^{13}\Mh$. Finally, haloes in isotropic environments (small $\alpha$) are substantially less clustered than those in anisotropic environments.

The assembly bias trend with halo concentration (as well as formation time, which we don't show here) has been widely discussed in the literature \citep[see, e.g.,][]{wechsler+06,jsm07,dwbs08,desjacques:2008a,abl08,fw10,shpl16,lms17,pp17}. The inversion of the trend is related to the tidal anisotropy of the halo environment; a large fraction of low-mass haloes live in highly anisotropic and biased environments\footnote{We discuss the so-called `splashback' haloes in section~\ref{sec:splashbackmergers}.} such as cosmic filaments, unlike more isolated haloes which dominate their environment and follow the trends predicted by standard spherical collapse models \citep{phs18a}. There are also indications that the trend in velocity anisotropy $\beta$ may be connected to the tidal environment, with low-mass haloes accreting in filaments being dominated by tangential orbits; such haloes should inherit high values of large-scale bias from their parent filaments \citep{bprg17}. 

The monotonic dependence of halo bias on halo asphericity $c/a$ and spin $\lambda$ at fixed mass in the right panel of Figure~\ref{fig:assemblybias} is consistent with the trends noted previously in the literature using configuration space definitions of bias \citep{bett+07,gw07,fw10,johnson+19} \citep[see also][for a study of shape- and spin-dependent clustering at Mpc scales]{vDaw12}. 

As regards the asphericity of the velocity ellipsoid $c_v/a_v$ or related variables, we are unaware of any work other than \citet{fw10} that has discussed the corresponding assembly bias trend. It is therefore worth commenting on the nature of this trend before proceeding. We see in the left panel of Figure~\ref{fig:assemblybias} that the amplitude of the trend with $c_v/a_v$ is only slightly weaker than that with $\beta$. The nature of the trend is quite interesting, however, since it says that haloes with spherical velocity ellipsoids cluster less strongly than aspherical ones. On the one hand, this suggests a potential connection with the trend shown by the asphericity of the shape tensor $c/a$ which is qualitatively identical. On the other, it is also tempting to compare with the trend due to $\beta$. Keeping in mind that perfectly spherical velocity ellipsoids would correspond to $\beta=0$, it is clear that the trend defined by upper and lower quartiles of $\beta$ is actually sensitive to additional information about haloes with aspherical velocity ellipsoids, by splitting these into radially dominated (upper $\beta$ quartile) and tangentially dominated (lower $\beta$ quartile) haloes (c.f. the discussion earlier regarding the connection between $\beta$ and the full phase space of the halo.)

It is clear from Figure~\ref{fig:assemblybias} that the trend between halo bias $b_1$ and the local tidal anisotropy $\alpha$ is the strongest amongst all the secondary bias trends. In fact, defining $\alpha$ at $\sim4\,\times\,$the halo radius ensures that this correlation is stronger than that between $b_1$ and the local overdensity $\delta$ of the halo environment measured at the same scale \citep{phs18a}. Moreover, the definition of $\alpha$ is such that this variable would be statistically independent of the very large-scale overdensity in the (Gaussian random) initial conditions, unlike $\delta$ at the same scale \citep{st02}. The fact that $\alpha$ and $b_1$ correlate so strongly is then highly suggestive of a physical link between these quantities related to the non-linear dynamics of halo formation \citep[see also][]{cphs17}. The strength of the $b_1\leftrightarrow\alpha$ correlation will be important below.

\begin{figure*}
\centering
\includegraphics[width=0.98\textwidth]{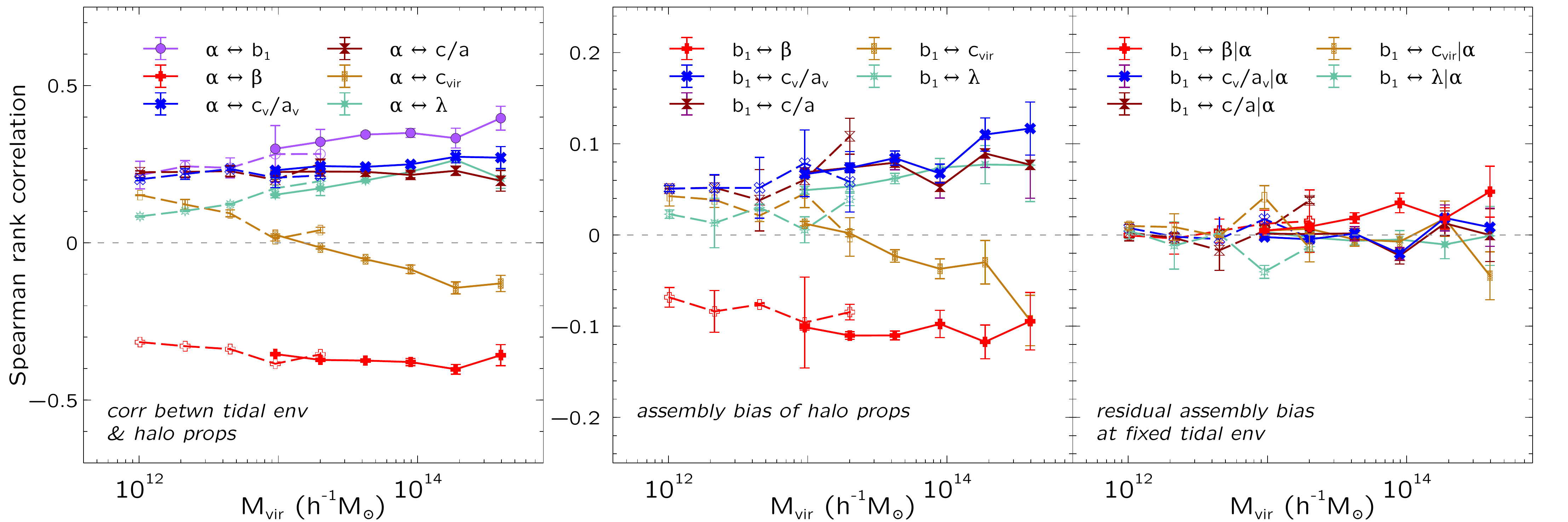}
\caption{
{\bf Correlations between internal halo properties, tidal environment and large-scale bias.}
\emph{(Left panel:)} Spearman rank correlation coefficients, for haloes in bins of mass $M_{\rm vir}$, between tidal anisotropy $\alpha$ and other halo properties, including $\gamma_{\alpha b_1}$ with large-scale bias $b_1$ and $\gamma_{\alpha c}$ with internal properties $c\in\{\beta,c_v/a_v,c/a,c_{\rm vir},\lambda\}$ (see caption of Figure~\ref{fig:assemblybias}). In the legend, each coefficient $\gamma_{ab}$ is represented by the symbol $a\leftrightarrow b$.
\emph{(Middle panel:)} Assembly bias trends seen using Spearman rank correlation coefficients $\gamma_{b_1c}$ between halo bias and each internal property $c$ (c.f. Figure~\ref{fig:assemblybias}).
\emph{(Right panel:)} Conditional correlation coefficients $\gamma_{b_1c|\alpha}$ (equation~\ref{eq:condcov}) for each internal property $c$.
Note that the vertical axis in the middle and right panels is zoomed in by a factor $\sim3$ as compared to the left panel.
The formatting of symbols (filled versus empty) and lines (solid versus dashed) is identical to that in Figure~\ref{fig:assemblybias}.
\emph{The right panel shows the main result of this work:} each conditional coefficient $\gamma_{b_1c|\alpha}$ is substantially smaller in magnitude than the corresponding unconditional coefficient $\gamma_{b_1c}$ in the middle panel.  Thus, conditioning on tidal anisotropy $\alpha$ largely accounts for the assembly bias trend of \emph{all} internal halo properties. See text for a discussion.
}
\label{fig:corr-main}
\end{figure*}

\subsection{Disentangling multi-scale correlations using conditional correlation coefficients}
\label{subsec:corrcoeff}
\noindent
As discussed in the Introduction, we are interested in identifying a clean statistical signature that contributions from different length scales might segregate into distinct correlations: one between internal halo properties and the local cosmic web environment and the other between the local web and large-scale halo bias.
A convenient approach to addressing this issue is to use the concept of conditional correlation coefficients \citep{han+19}, as we describe next. This analysis is made possible by our use of a halo-by-halo measurement of bias that does not require haloes to be binned. 

Consider three standardized (i.e., zero mean, unit variance) \emph{Gaussian} variables $a,b,c$ with mutual correlation coefficients $\gamma_{ab}$, $\gamma_{bc}$ and $\gamma_{ca}$. The conditional distribution $p(b,c|a)$ is then a bivariate Gaussian with variances $\textrm{Var}(b|a)=1-\gamma_{ab}^2$, $\textrm{Var}(c|a)=1-\gamma_{ac}^2$ and the conditional covariance
\be
 \textrm{Cov}(b,c|a)=\gamma_{bc}-\gamma_{ab}\gamma_{ac} \equiv \gamma_{bc|a}\,.
\label{eq:condcov}
\ee
\emph{The key point to note is that, if $\gamma_{bc|a}=0$, then the conditional distributions of $b$ and $c$ at fixed $a$ are independent: $p(b,c|a) = p(b|a)p(c|a)$.} Bayes' theorem then implies that \emph{the conditional distribution of $c$ is independent of $b$:} $p(c|a,b) = p(c|a)$. In the present context, to the extent that any statistical correlation between physical variables should ultimately have a physical origin, this would strongly suggest that the statistical connection between $c$ and $b$ is linked by (at least) two physical mechanisms, one connecting $c$ to $a$ and the other connecting $a$ to $b$.

This discussion shows that the vanishing of $\gamma_{bc|a}=\gamma_{bc}-\gamma_{ab}\gamma_{ac}$ is a useful diagnostic of the conditional independence of $c$ on $b$.  Although we phrased the discussion in terms of a multi-variate Gaussian for $p(a,b,c)$, the fact that this distribution is non-Gaussian is not as large a concern as one might have imagined.  Rather, the significance of $\gamma_{bc|a}=0$ is tied to the assumption that $c$ can be well-approximated by a model which is linear in $a$ and $b$ \citep[see equations~3 and~4 in][]{bernardi+03}.
It is just that, for a multi-variate Gaussian, the linear model is exact. 

Nevertheless, to minimise systematic errors, we will rely on measurements of Spearman's rank correlation coefficients for each pair of variables, which standardises all the distributions before computing correlations. Below, we will also discuss tests of the robustness of this choice of statistics.

\subsection{Tidal anisotropy as an indicator of assembly bias}
\label{subsec:anisoassemblybias}
\noindent
Our motivation behind setting up the correlation analysis in the previous section was to explore the possibility that assembly bias correlations between internal halo properties and large-scale bias might be explained using the separate correlations of each of these with some intermediate-scale environmental variable. In this context, it is worth mentioning that previous investigations of assembly bias have failed to identify any single environmental variable that might be responsible for correlations between halo bias and multiple internal halo properties \citep{villarreal+17,xz18}. The fact that tidal anisotropy $\alpha$ shows by far the strongest correlation with halo bias makes $\alpha$ a promising candidate for such a variable.

In the language of the previous section, therefore, we will now think of $a$ as the tidal anisotropy $\alpha$, $b$ as halo bias $b_1$ and $c$ as any one of the internal halo properties $\{\beta,c_v/a_v,c/a,c_{\rm vir},\lambda\}$. Below we will also report the results of analysing other permutations and combinations of variables, including using intermediate-scale overdensity $\delta$ as the environmental variable.

\emph{Figure~\ref{fig:corr-main} shows the main results of this paper.}
The \emph{left panel} shows Spearman rank correlation coefficients (for haloes in fixed bins of $M_{\rm vir}$)\footnote{We have checked that all our results are robust to our choice of binning. Namely, we found identical results for all correlation trends when doubling the number of mass bins. Thus our results are unaffected by mass-dependent trends in any correlation.} between the tidal anisotropy $\alpha$ and other halo properties including halo bias $b_1$ and all internal properties $c\in\{\beta,c_v/a_v,c/a,c_{\rm vir},\lambda\}$. This panel summarizes a number of previously known results, including the observations that, at fixed mass, haloes in more anisotropic tidal environments tend to be more strongly clustered \citep[$\alpha\leftrightarrow b_1$,][]{hahn+09,phs18a}, more concentrated \citep[$\alpha\leftrightarrow c_{\rm vir}$,][]{phs18a}, more spherical \citep[$\alpha\leftrightarrow c/a$,][]{wang+11}, with higher spin \citep[$\alpha\leftrightarrow \lambda$,][]{hahn+09,wang+11}, and have more tangentially dominated velocity distributions \citep[$\alpha\leftrightarrow \beta$,][]{bprg17}. 
Additionally, we see that objects in anisotropic environments also have more spherical velocity ellipsoids ($\alpha\leftrightarrow c_v/a_v$), with a correlation very similar at all masses to that between $\alpha$ and the mass ellipsoid asphericity $c/a$. 

The \emph{middle panel} of Figure~\ref{fig:corr-main} summarizes the known assembly bias trends discussed in section~\ref{subsec:knownresults}. We see that the strength and sign of the correlation coeffcients at any halo mass is perfectly consistent with the results of the previous binned analysis (Figure~\ref{fig:assemblybias}) which focused on the extremes of the distributions of internal halo properties. Note that we have zoomed in on the vertical axis as compared to the left panel; the correlations of halo properties with large-scale bias are weaker (by approximately a factor $\sim3$ in each case) than the respective correlations with the local tidal environment.

The \emph{right panel} of Figure~\ref{fig:corr-main} shows our main new result: we display the conditional correlation coefficients $\gamma_{b_1c|\alpha}$ (calculated using equation~\ref{eq:condcov}) for each internal property $c\in\{\beta,c_v/a_v,c/a,c_{\rm vir},\lambda\}$. The vertical scale is identical to that in the middle panel which showed the corresponding unconditional coefficients using the same scheme for colours and markers. In each case, we see that the conditional coefficients are substantially smaller in magnitude than the corresponding unconditional ones at all masses (by a factor $\sim 4$ or so at low masses). In fact, except for $\beta$ around $\sim10^{14}\Mh$ (see below), the conditional coefficients are scattered around zero in all cases over the entire mass range, implying that $\alpha$ is an excellent candidate for the primary environmental variable responsible for halo assembly bias trends. 
In support of this argument, we find in Appendix~\ref{app:explicitcondcorr} (see also below) that conditioning on $\alpha$ performs much better at decreasing assembly bias strength than conditioning on $\delta$ at the same scale, despite the fact that $\alpha$ and $\delta$ are correlated. Further, in order to quantify exactly how close to zero the conditional coefficients $\gamma_{b_1c|\alpha}$ are in Figure~\ref{fig:corr-main}, we use two methods. 

\begin{table}
\centering
\begin{tabular}{ll}
\hline\hline
conditional corr. coeff. & $\chi^2/$dof \\
\hline\hline
$b_{1}\leftrightarrow \lambda\,|\,\alpha$     & 0.90 \\
$b_{1}\leftrightarrow c/a\,|\,\alpha$          & 0.97 \\
$b_{1}\leftrightarrow c_{v}/a_{v}\,|\,\alpha$   & 1.76 \\         
$b_{1}\leftrightarrow c_{vir}\,|\,\alpha$            & 2.18 \\              
$b_{1}\leftrightarrow \lambda\,|\,\beta$       & 3.13 \\
$b_{1}\leftrightarrow \beta\,|\,\alpha$          & 4.34 \\  
\hline
$b_{1}\leftrightarrow c_{\rm vir}\,|\,\delta$           & 11.89 \\   
$b_{1}\leftrightarrow c_{\rm vir}\,|\,c_{v}/a_{v}$     & 14.52 \\   
$b_{1}\leftrightarrow c_{\rm vir}\,|\,c/a$     & 24.00\\      
$b_{1}\leftrightarrow c/a\,|\,c_v/a_v$     & 29.96\\      
\hline\hline
\end{tabular}
\caption{Top 10 conditional correlation coefficients $b_1\leftrightarrow X\,|\,Y$ rank-ordered by reduced Chi-squared values for comparison to zero. Here $X,Y$ were allowed to be any two of the variables $\{\beta,c_v/a_v,c/a,c_{\rm vir},\lambda,\alpha,\delta\}$, i.e., treating environmental variables on par with internal halo properties. Chi-squared values were calculated using measurements in $9$ mass bins, with mean values and errors computed using 2 realisations of the high resolution and 10 realisations of the low resolution simulations. The \emph{first column} labels the conditional coefficient being tested and the \emph{second column} reports the value of reduced Chi-squared for $9$ degrees of freedom. Values below the horizontal line correspond to $p$-values $<10^{-4}$. }
\label{tab:chisqd}
\end{table}

The first is a straightforward Chi-squared test which we perform using the results in 9 mass bins of the low and high resolution boxes where we have reliable error bars on the measurements of $\gamma_{bc|a}$. While performing this test, we also relaxed the assumption of using $\alpha$ as the intermediary between halo properties and bias, exploring multiple other combinations involving either the overdensity $\delta$ or one of the internal properties themselves as the intermediary. When the resulting triplet combinations are ordered in increasing order of reduced Chi-squared, we find that triplets involving $\alpha$ as the intermediary produce the best Chi-squared values, while those involving $\delta$ perform much worse. Table~\ref{tab:chisqd} summarizes these results. We note that the largest discrepancy in our conclusions occurs for the assembly bias variable $\beta$ at high masses; the residual assembly bias conditioned on $\alpha$ deviates from zero at masses $>8\times10^{13}\Mh$. This can be  seen in right panel of Figure~\ref{fig:corr-main} and also causes the largest Chi-squared values out of the five assembly bias variables in Table~\ref{tab:chisqd}.

\begin{figure}
\centering
\includegraphics[width=0.45\textwidth]{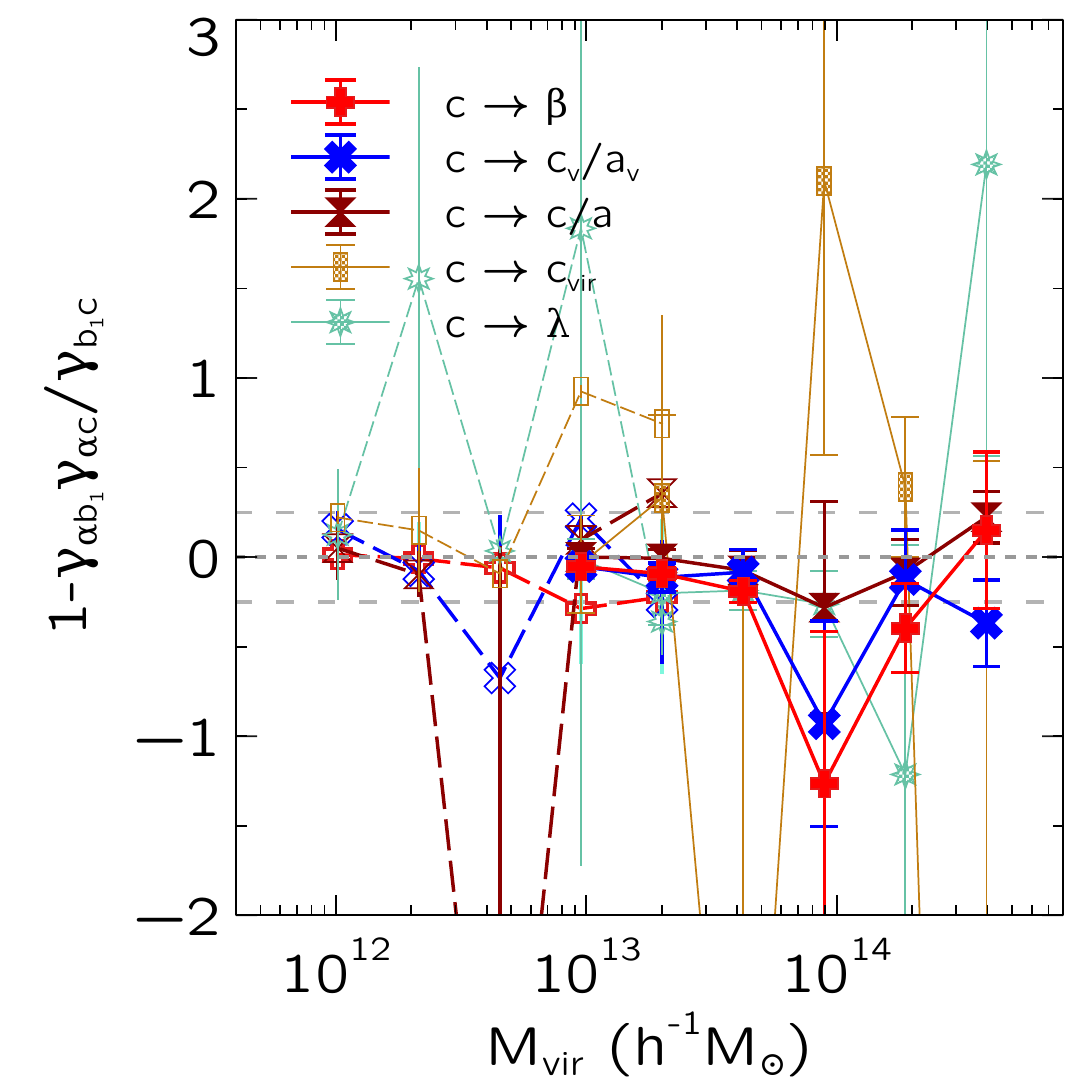}
\caption{{\bf Relative correlation coefficients}, calculated as the ratio $\gamma_{b_1c|\alpha}/\gamma_{b_1c}$ using measurements from the middle and right panels of Figure~\ref{fig:corr-main} (and formatted identically) for internal halo properties $c\in\{\beta,c_v/a_v,c/a,c_{\rm vir},\lambda\}$ as indicated. The horizontal dotted line indicates zero and the horizontal dashed lines indicate $\pm0.25$, i.e., a factor $4$ decrease in the magnitude of $\gamma_{b_1c|\alpha}$ relative to $\gamma_{b_1c}$. See text for a discussion.}
\label{fig:corr-ratio}
\end{figure}

The second method is to simply construct the ratio $\gamma_{bc|a}/\gamma_{bc}$: if the magnitude of this ratio is small, it means that conditioning on $a$ has indeed substantially decreased the correlation between $b$ and $c$. This is a particularly useful diagnostic for internal properties such as halo concentration whose correlation with halo bias is the smallest in amplitude of all internal properties. The results are shown in Figure~\ref{fig:corr-ratio}. For the internal properties $\{\beta,c_v/a_v,c/a\}$, we see that the relative correlation coefficient is, in fact, much smaller than unity over nearly the entire halo mass range. 

For halo concentration and spin, on the other hand, the relative correlation shows very large fluctuations and noise at higher masses. This is perhaps not surprising considering previous results which suggest that assembly bias signatures at these mass scales are likely caused by other effects \citep{dwbs08,phs18a}. Interestingly, our results from Table~\ref{tab:chisqd} and Figure~\ref{fig:corr-ratio} indicate that $\alpha$ is a particularly good indicator of spin assembly bias in the mass range $\sim10^{12}$-$10^{14}\Mh$. 
This can be compared with the results of \citet{johnson+19} who found that spin assembly bias can be largely explained using the presence of neighbours of comparable mass. Our results are consistent with theirs, since $\alpha$ represents the anisotropy of the \emph{total} tidal field in the halo vicinity, including the influence of all neighbours.

\emph{To summarize, the statistical correlation between large-scale bias $b_1$ and essentially any internal halo property $c$ that we have studied is consistent with arising from the individual correlations $b_1\leftrightarrow\alpha$ and $\alpha\leftrightarrow c$, at nearly all halo masses.} 

\subsection{Reliability of chosen statistics}
\label{subsec:robustness}
\noindent
We argued in section~\ref{subsec:corrcoeff} that the use of correlation coefficients combined using \eqn{eq:condcov} relies essentially on the implicit assumption that the underlying correlations between triplets of variables are linear. Our use of Spearman's rank correlations means that the relevant variables are actually the \emph{ranks} of the physical variables, so that we are dealing with triplets of correlated variables which are individually uniformly distributed.
Although the variables are now standardized, their intrinsic correlations are not necessarily linear or even monotonic \citep[see, e.g., Figure~12 of][which shows that the median halo concentration is non-monotonic in $\alpha$ at fixed mass]{phs18a}, so one might still worry about systematic effects in our analysis. We have therefore performed some explicit tests, which we describe here, to establish the robustness of our conclusions. Our method differs from that of \citet{han+19} who used Gaussian process regression to explicitly fit for the non-linearity / non-monotonicity of the dependence of halo bias on other variables, thus allowing them to explore a multi-dimensional bias `manifold'. Instead, below we demonstrate the robustness of our primary results using direct probes of probability distributions involving $b_1$, $\alpha$ and one halo internal property at a time.

We first test the reliability of replacing explicit conditional correlation coefficients (which would require binning of data) with the expression in \eqn{eq:condcov} (which uses all available data) in Appendix~\ref{app:explicitcondcorr}, focusing on the strongest assembly bias signature which is that of the velocity anisotropy $\beta$. Figure~\ref{fig:corr-betab1-deltaalpha} shows that explicitly binning in $\alpha$ before computing the correlation coefficient between $b_1$ and $\beta$ does decrease the magnitude of the correlation to nearly zero at all masses and for all $\alpha$.

\begin{figure*}
\centering
\includegraphics[width=0.98\textwidth]{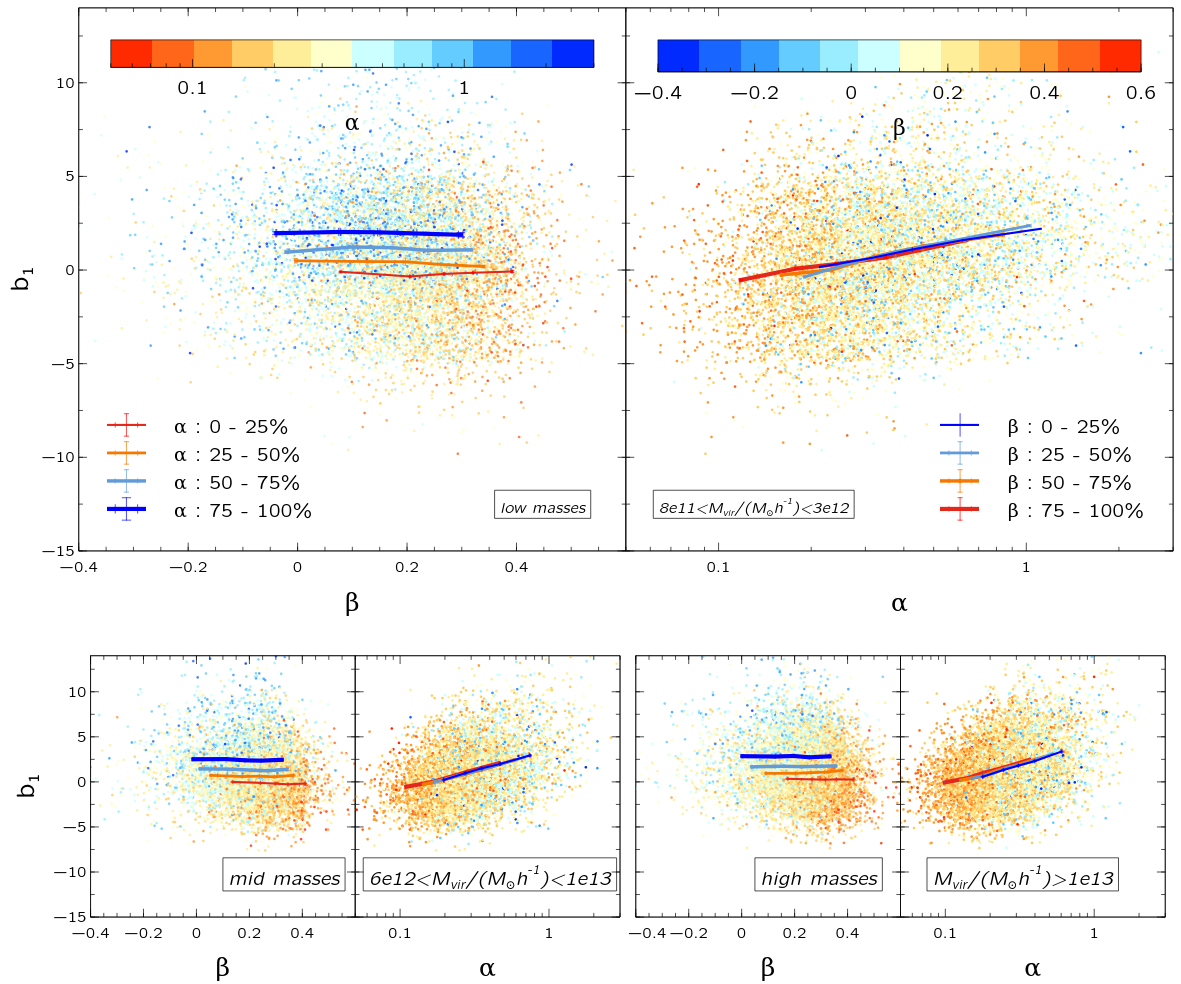}
\caption{{\bf Joint distribution of $\alpha$, $b_1$ and $\beta$ for haloes} in three mass ranges, low [$8\times10^{11}<M_{\rm vir}/(\Mh)<3\times10^{12}$], mid [$6\times10^{12}<M_{\rm vir}/(\Mh)<10^{13}$] and high [$M_{\rm vir}/(\Mh)>10^{13}$]. 
\emph{(Top Left panel:)} Scatter plot shows $\beta$ against $b_1$ with points coloured by $\alpha$. Each coloured solid line focuses on a quartile of $\alpha$ as indicated in the legend, showing the median $b_1$ in bins of $\beta$ (the bins are chosen to be quintiles of $\beta$ for haloes in each $\alpha$ quartile).
\emph{(Top Right panel:)} Scatter plot shows $\alpha$ against $b_1$ with points coloured by $\beta$. Similarly to the left panel, each coloured solid line now shows the median $b_1$ in quintiles of $\alpha$, for haloes selected in a quartile of $\beta$ as indicated. 
The results of the two panels are consistent with a correlation structure $p(\alpha,b_1,\beta)\simeq p(\alpha)p(b_1|\alpha)p(\beta|\alpha)$.
See text for a discussion. The bottom panels show the same as the top for mid and high masses.}
\label{fig:dist-betab1alpha}
\end{figure*}

To address the concern regarding non-linearity or non-monotonicity of the intrinsic correlations, we focus on three mass ranges; low [$8\times10^{11}<M_{\rm vir}/(\Mh)<3\times10^{12}$], mid [$6\times10^{12}<M_{\rm vir}/(\Mh)<10^{13}$] and high [$M_{\rm vir}>10^{13}\Mh$] containing $\sim10^4$ haloes each and dissect the full distribution of $\{b_1,\alpha,\beta\}$ in Figure~\ref{fig:dist-betab1alpha}. 
The scatter plots in the \emph{top panels} of the Figure focus on the low mass range. The \emph{top left panel} shows the distribution of $\beta$ and $b_1$, with the symbols coloured by the value of $\alpha$. Apart from an overall negative correlation between $\beta$ and $b_1$ (c.f. middle panel of Figure~\ref{fig:corr-main}), we can also see that both these variables are correlated with $\alpha$ by observing that the redder (bluer) points, which correspond to $\alpha\lesssim0.1$ ($\alpha\gtrsim1$) are largely confined to the bottom right (top left) of the distribution. Similar conclusions about the correlation between variables can be made from the \emph{top right panel} which shows the scatter distribution of $\alpha$ and $b_1$, with the symbols coloured by the value of $\beta$.
 
In order to extract more information on the structure of the joint probability distribution $p(\alpha,b_1,\beta)$, we consider the mean value of bias conditioned on $\alpha$ and $\beta$ i.e, $\avg{b_1|\alpha,\beta}$. In the 3-dimensional space of $\{b_{1},\alpha,\beta\}$, this quantity forms a 2-dimensional surface whose properties we explore using projections onto the $b_1$-$\beta$ and $b_1$-$\alpha$ planes, as we discuss next.

In the \emph{top left panel} of Figure~\ref{fig:dist-betab1alpha}, we plot the projection of $\avg{b_1|\alpha,\beta}$ onto the $b_1$-$\beta$ plane as solid lines, with each line focusing on haloes in quartiles of $\alpha$ (from red to blue in increasing thickness as $\alpha$ increases.) The overall assembly bias trend between $b_1$ and $\beta$ is now visible as the fact that the blue (red) curve having larger (smaller) bias lies toward smaller (larger) $\beta$. More interestingly, we see that each of these lines is approximately horizontal; \emph{this implies that $\avg{b_1|\alpha,\beta}$ in each quartile of $\alpha$ is independent of $\beta$}, i.e, $\avg{b_1|\alpha,\beta}\simeq\avg{b_1|\alpha}$. In other words, bias when conditioned on $\alpha$ does not show an assemby bias with $\beta$. 

Similarly, the projection of $\avg{b_1|\alpha,\beta}$ onto the $b_1$-$\alpha$ plane is shown in the \emph{top right panel} as solid lines, with each line focusing on haloes in quartiles of $\beta$ (from blue to red in increasing thickness as $\beta$ increases). All the lines clearly trace out the same locus of positive correlation between $b_1$ and $\alpha$, with vertical and horizontal shifts now occurring essentially in perfect tandem as $\beta$ changes. A simple calculation shows that this is again consistent with the relation $\avg{b_1|\alpha,\beta}\simeq\avg{b_1|\alpha}$.

\begin{figure*}
\centering
\includegraphics[width=0.98\textwidth]{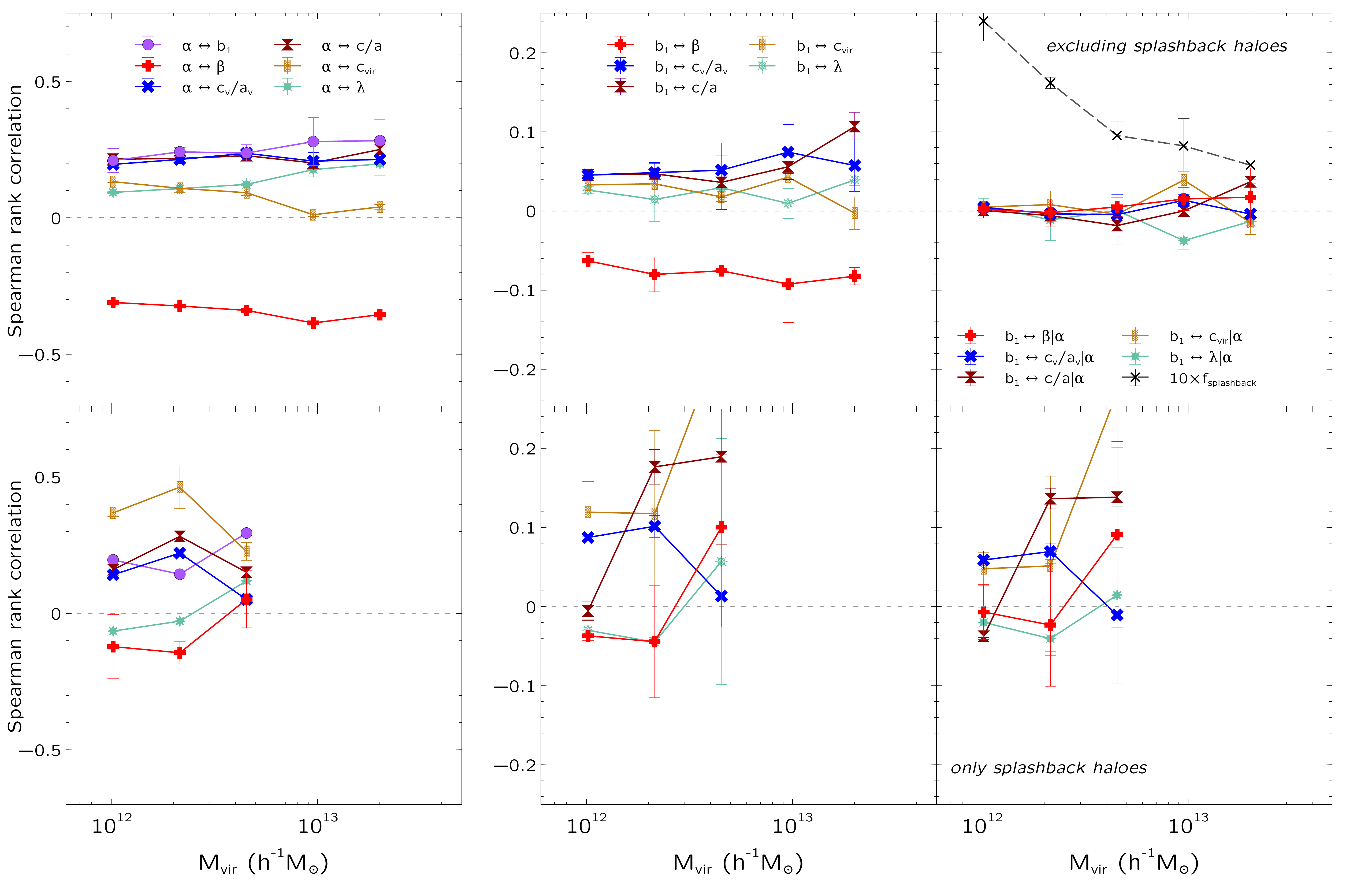}
\caption{Same as Figure~\ref{fig:corr-main}, showing results of the {\bf analysis performed separately for splashback and other haloes}, restricted to the higher resolution boxes and hence lower masses. Splashback objects were identified as described in section~\ref{subsec:splashback}.
\emph{(Top row:)} Results \emph{excluding splashback objects}; these are essentially identical to those in Figure~\ref{fig:corr-main}, with $\alpha$ being a good indicator of all assembly bias trends.
Black dashed curve in the top right panel shows $10\,\times\,$the fraction of haloes classified as splashback objects.
\emph{(Bottom row:)} Results for \emph{splashback objects only}; in this case, $\alpha$ is a very poor indicator of \emph{any} assembly bias trend. 
Results are not displayed for the two highest mass bins which contain fewer than $20$ objects each.
See text for a discussion.}
\label{fig:corr-splashback}
\end{figure*}  

These results together give a complete picture of $\avg{b_1|\alpha,\beta}$ as being approximately a plane in $\{b_1,\log \alpha,\beta\}$ space and moreover being orthogonal to the $b_1$-$\log \alpha$ plane. The \emph{bottom panels} of Figure~\ref{fig:dist-betab1alpha} show that identical conclusions can be drawn for the mid \emph{(bottom left)} and low \emph{(bottom right)} mass bins. We can go even further and ask whether the conditional \emph{variance} ${\rm Var}(b_1|\alpha,\beta)$ also displays the same behaviour. Figure~\ref{fig:bias_variance} in the Appendix shows that this is indeed the case: the projections of this quantity in the $b_1$-$\beta$ and $b_1$-$\alpha$ planes are consistent with the relation ${\rm Var}(b_1|\alpha,\beta)\simeq{\rm Var}(b_1|\alpha)$.

These results strongly suggest that the joint distribution $p(\alpha,b_1,\beta)$ itself (as opposed to only its first moment) has a structure consistent with $b_1$ and $\beta$ being conditionally independent of each other, when conditioned on $\alpha$:
\be
p(\alpha,b_1,\beta)\simeq p(\alpha)p(b_1|\alpha)p(\beta|\alpha)\,,
\notag
\ee
which then ensures that the trends of $\avg{b_1|\alpha,\beta}$ and ${\rm Var}(b_1|\alpha,\beta)$ discussed above emerge. Thus, the overall anti-correlation between $b_1$ and $\beta$ (assembly bias) is explained by the mutual dependence of these variables on the tidal anisotropy $\alpha$.

We emphasize that this analysis made \emph{no assumptions} regarding Gaussianity of the variables, monotonicity or linearity of the trends, etc. We have further verified that essentially identical results are obtained using \emph{all} other internal halo properties considered in this work as well (see Figures~\ref{fig:bias_mean} and~\ref{fig:bias_variance} in the Appendix). 

The results of this section therefore provide strong support for our claim that the tidal anisotropy $\alpha$ is the primary indicator of assembly bias for a number of internal halo properties. Our tests have further demonstrated that our conclusions are robust to our choice of statistical tools (Spearman rank correlation statistics, with conditional correlation coefficients defined by equation~\ref{eq:condcov}).  In the next section, we explore other, physical choices related to sample selection which could, in principle, affect our conclusions.

\begin{figure*}
\centering
\includegraphics[width=0.98\textwidth]{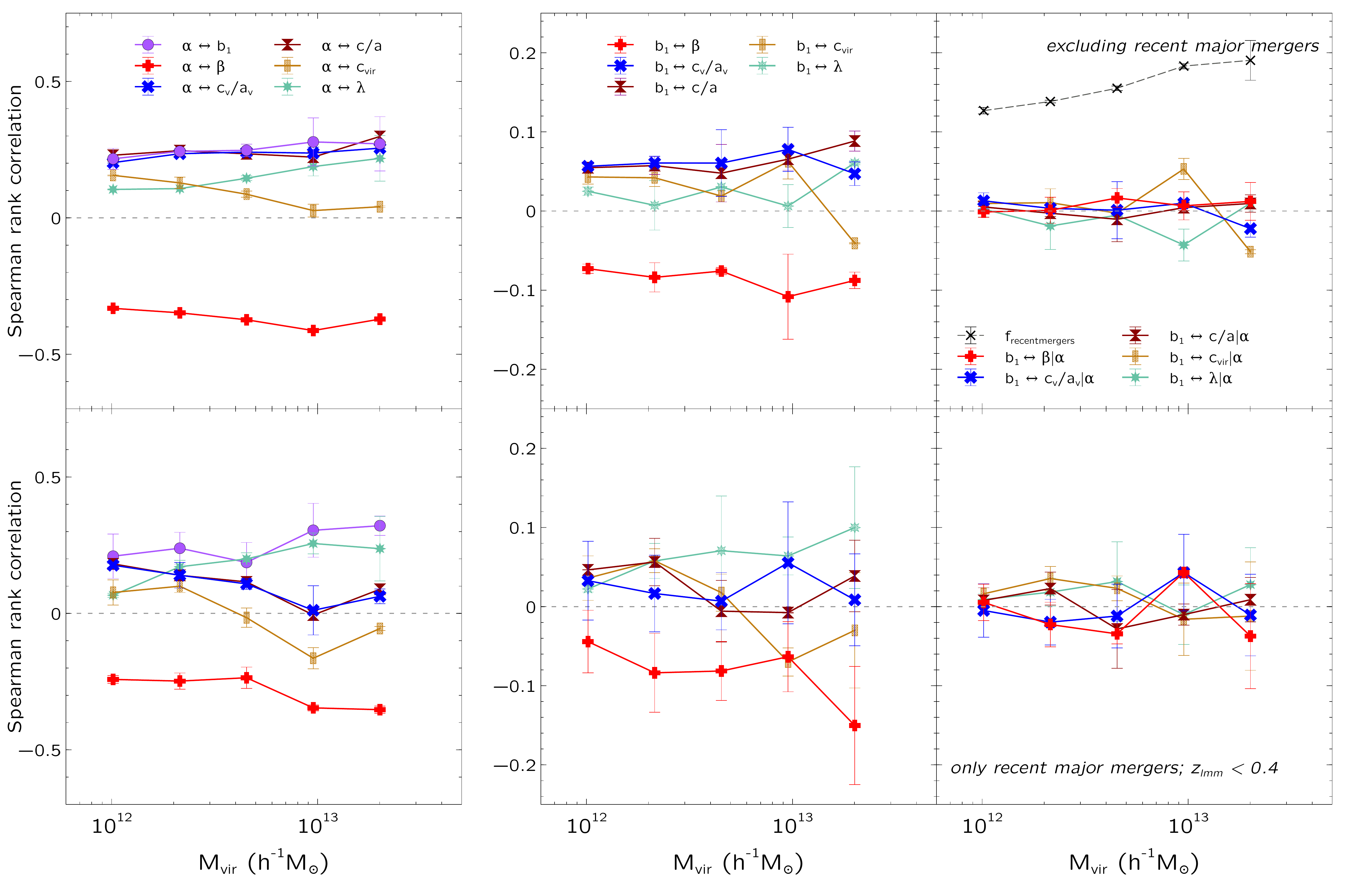}
\caption{Same as Figure~\ref{fig:corr-splashback}, now {\bf splitting haloes into those with recent major merger events and other haloes}. The haloes with recent major mergers were required to have their last major merger at a redshift $z_{\rm lmm} < 0.4$ (see section~\ref{subsec:mergers} for details).
\emph{(Top row:)} Results \emph{excluding haloes with recent major mergers}; these are essentially identical to those in Figure~\ref{fig:corr-main}, with $\alpha$ being a good indicator of all assembly bias trends.
Black dashed curve in the top right panel shows the fraction of haloes classified as recent major mergers.
\emph{(Bottom row:)} Results for \emph{only haloes with recent major mergers}; again, $\alpha$ is still a good indicator of all assembly bias trends. We conclude that $\alpha$ influences assembly bias similarly for both these populations.
See text for a discussion.}
\label{fig:lastmm}
\end{figure*}  
  
\section{The impact of splashback objects and major mergers}
\label{sec:splashbackmergers}
\noindent
The primary analysis of this work presented in section~\ref{sec:correlns} defined haloes as objects identified as being distinct at the epoch of interest $z=0$. These haloes therefore also include the small population of so-called `splashback' haloes \citep{gkg05}, which are objects that have passed through one pericenter passage of their eventual host but are currently outside its virial radius. Treating splashback objects equivalently to genuine distinct haloes therefore risks contaminating any signal that involves a correlation with large-scale environment. Indeed, there is considerable evidence that, at low masses, a significant fraction of the assembly bias signal in variables such as halo concentration or age in fact arises from splashback objects \citep{dwbs08,hahn+09,shpl16,villarreal+17,mk19}. It is then important to assess the impact of this small population on our conclusions regarding the influence of the cosmic web environment.

Similarly, the fact that there are strong correlations between tidal environment and internal properties such as halo asphericity in position or velocity space could be connected to the occurrence of recent major merger events. We must therefore also ask whether the cancellations we see in the conditional correlation coefficients in the previous section are related to major mergers. 

We address both of these issues in this section, showing that our results are unchanged when excluding splashback haloes or segregating haloes by the epoch of their last major merger.

\subsection{Splashback objects}
\label{subsec:splashback}
\noindent
We identify splashback haloes using the output of \textsc{consistent-trees} which provides the redshift $z_{\rm firstacc}$ of the `first accretion' event of each object. This is the epoch at which the main progenitor of the object first passed inside the virial radius of a larger object. Splashback haloes are then objects which are currently not identified as subhaloes (i.e., not inside the virial radius of a larger object; `PID'$=-1$ according to \textsc{consistent-trees}) but have  $z_{\rm firstacc}>0$. With a fine time resolution in our merger tree which uses $201$ snapshots, we expect this criterion to capture most of these objects.

We have repeated the analysis of section~\ref{sec:correlns} for halo samples excluding splashback haloes and also for the splashback haloes themselves. Since we only have merger histories available for haloes in our high resolution boxes, we focus on the low-mass range for this analysis. Figure~\ref{fig:corr-splashback} shows the results. We see in the \emph{top row} that \emph{excluding splashback haloes has essentially no impact on our main results}, since the correlation coefficients in the left and middle panels, as well as the level of cancellation in the right panel, are nearly identical to the low-mass results of Figure~\ref{fig:corr-main}. 
The black curve in the top right panel shows the fraction of haloes that were excluded as being splashback objects; this is always $\lesssim2\%$ over this mass range and decreases as expected towards higher masses.

Interestingly, when we repeat the analysis for these splashback objects themselves (\emph{bottom row} of Figure~\ref{fig:corr-splashback}), we see very different behaviour. 
Firstly, the correlation between $\alpha$ and $b_1$ at the lowest masses is now weaker in magnitude than other correlations, in stark contrast to the case for distinct haloes.
And the right panel shows that, in fact, $\alpha$ has essentially no impact on the assembly bias correlations involving any internal property. 
(We do not display the results for the two highest mass bins which contain fewer than $20$ objects each.)
In other words, \emph{the cosmic web anisotropy is a very poor indicator of any assembly bias trend for splashback objects}. This is physically perhaps not surprising considering the very different accretion and tidal stripping histories of these objects as compared to other genuinely distinct haloes. We discuss this further in section~\ref{sec:conclude}.

\subsection{Recent major mergers}
\label{subsec:mergers}
\noindent
The output of \textsc{consistent-trees} provides, for each object, the epoch of the last major merger event this object experienced on its main progenitor branch. The definition of a major merger is an event involving the overlap of virial radii of objects with a mass ratio closer to unity than $1:3$. We use objects from the higher resolution box as in the earlier analysis and \emph{discard splashback objects} as defined by the criterion of section~\ref{subsec:splashback}. We segregate the remaining objects by their redshift $z_{\rm lmm}$ of last major merger into two populations : those with recent major mergers which occurred at $z_{\rm lmm}<0.4$ (corresponding to $< 4.3$ Gyr of lookback time) and those with major mergers further back in the past. We then compute the same rank correlation coefficients as before and repeat the analysis similar to that shown in Figure~\ref{fig:corr-splashback}.

Figure~\ref{fig:lastmm} shows the results. Our segregation makes the recent major merger population have fewer objects, thus the results have larger scatter. Despite this, we see that separating out the population with recent major mergers does not bring out any dramatic difference in our main results, suggesting that both the populations of haloes are influenced similarly by their respective tidal environment as regards their assembly bias trends. \emph{We conclude that major merger events are not a likely cause for $\alpha$ being an excellent statistical intermediary in explaining halo assembly bias.}

\section{Discussion \& conclusion}
\label{sec:conclude}
\noindent
The hierarchical formation of cosmological structure leads to distinct connections between the properties of the cosmic web and its constituent dark matter haloes across a wide range of length scales. The most striking amongst these are the ones categorized as assembly bias (or secondary bias), in which the large-scale ($\gtrsim\textrm{few}\times10\Mpch$) clustering strength of haloes shows distinct trends with a number of internal halo properties (defined at scales $\lesssim R_{\rm vir}\sim\textrm{few}\times100\kpch$), even at fixed halo mass. Understanding the origin of such correlations across several orders of magnitude in length scale is of great interest from the point of view of building a complete understanding of structure formation in the $\Lambda$CDM framework, and can have consequences for galaxy evolution and precision cosmology.

In this work, we have explored the idea that many (if not all) assembly bias trends in the mass range $8\times10^{11}\Mh\lesssim M_{\rm vir}\lesssim5\times10^{14}\Mh$ could be largely a result of a multi-scale connection between internal halo properties and the large-scale environment, with the \emph{local, non-linear cosmic web} environment acting as an intermediary. This is motivated by the expectation that these correlations must be connected to the only physical mechanism at play (gravitational tides) at the most natural intermediate length scale in the problem (the current turn-around radius for infalling material around a given halo, which is close to $\sim4\,\times\,$the halo radius).

We considered \emph{scalar internal properties} related to the shape, velocity dispersion, density profile and angular momentum of haloes; these include the halo shape asphericity $c/a$ (section~\ref{subsubsec:massellipsoid}), velocity ellipsoid asphericity $c_v/a_v$ (section~\ref{subsubsec:velocityellipsoid}), velocity anisotropy $\beta$ (section~\ref{subsubsec:beta}), concentration  $c_{\rm vir}$ (section~\ref{subsubsec:cvir}) and spin $\lambda$ (section~\ref{subsubsec:spin}). 
The \emph{large-scale environment} of each halo was characterised using the halo-by-halo bias $b_1$ of \citet{phs18a} defined at $\gtrsim30\Mpch$ scales (section~\ref{subsec:b1}) and, for the \emph{local cosmic web environment}, we considered the halo-centric tidal tensor defined at $\sim4R_{\rm 200b}$ scales (section~\ref{subsec:tidalmsrmnt}), focusing on the \emph{tidal anisotropy} variable $\alpha$ (equation~\ref{eq:alpha-def}) introduced by \citet{phs18a}.

Our primary statistical analysis relied on Spearman rank correlation coefficients calculated for pairs of variables. In particular, we argued that the \emph{vanishing of conditional correlation coefficients} defined in \eqn{eq:condcov} offers a useful and compact way to assess the strength of multi-variate statistical connections (section~\ref{subsec:corrcoeff}), and we further demonstrated that this technique is robust to all of the assumptions involved in using \eqn{eq:condcov} (section~\ref{subsec:robustness} and Appendices~\ref{app:explicitcondcorr} and~\ref{app:jointdistribution}). 

Our main results can be summarised as follows.
\begin{itemize}
\item 
The tidal anisotropy $\alpha$ shows the strongest correlation by far with $b_1$ at fixed halo mass amongst \emph{all} halo properties we have considered (Figure~\ref{fig:assemblybias} and \emph{middle panel} of Figure~\ref{fig:corr-main}) and correlates strongly with all internal halo properties as well (\emph{left panel} of Figure~\ref{fig:corr-main}). The correlation between $\alpha$ and $b_1$ in particular is substantially stronger than that between $b_1$ and pure density or pure anisotropy variables, as discussed in section~\ref{subsec:tidalmsrmnt}. The variable $\alpha$ is therefore an excellent candidate for an intermediary in explaining assembly bias, more so than the density contrast $\delta$ (equation~\ref{eq:delta-def}) defined at the same scale (Appendices~\ref{app:tidal-bias} and~\ref{app:tidal-internal}). 
\item 
The conditional correlation coefficients $\gamma_{b_1c|\alpha}$ are substantially smaller in magnitude than the unconditional coefficients $\gamma_{b_1c}$ for \emph{all} internal halo properties $c$ that we studied, for all but the highest mass scales we consider (\emph{right panel} of Figure~\ref{fig:corr-main}, see also Table~\ref{tab:chisqd} and Figure~\ref{fig:corr-ratio}). 
The joint distribution of $\alpha$, $b_1$ and any internal property $c\in\{\beta,c_v/a_v,c/a,c_{\rm vir},\lambda\}$ is therefore consistent with reflecting only two fundamental correlations $b_1\leftrightarrow\alpha$ and $c\leftrightarrow\alpha$:
\be
p(\alpha,b_1,c)\simeq p(\alpha)p(b_1|\alpha)p(c|\alpha)\,,
\label{eq:mainresult}
\ee
(section~\ref{subsec:corrcoeff}, see also Figure~\ref{fig:dist-betab1alpha} and Appendix~\ref{app:jointdistribution}). Thus, $\alpha$ indeed explains \emph{all} large-scale assembly bias trends, particularly at low halo mass. $\alpha$ defined at $\sim4\,\times\,$the halo radius also outperforms the environmental overdensity $\delta$ defined at fixed smoothing scales $1$-$2\Mpch$, recently proposed by \citet{han+19} as an assembly bias indicator (see Appendix~\ref{app:tidal-bias}).
\item Our conclusions regarding the role of $\alpha$ are \emph{unchanged upon excluding splashback haloes} from the analysis (section~\ref{subsec:splashback}, \emph{top row} of Figure~\ref{fig:corr-splashback}). Interestingly, repeating the analysis for the small population of splashback objects themselves (these are $\lesssim2\%$ of distinct haloes in our mass range) showed that $\alpha$ is a poor indicator of \emph{any} assembly bias trend for these objects (\emph{bottom row} of Figure~\ref{fig:corr-splashback}, see also below).
\item Our conclusions regarding $\alpha$ are also unchanged when segregating haloes by the presence or absence of a recent major merger event (section~\ref{subsec:mergers}, Figure~\ref{fig:lastmm}).
\end{itemize}

This wide-ranging effect of $\alpha$ in connecting small and large scales provides a new perspective on the phenomenon of assembly bias of low-mass haloes. 
There are several indications in the literature that multiple aspects of a halo's tidal environment could play a role in establishing the assembly bias trends of different variables. 
E.g., being in a non-linear filament affects the mass accretion rate and formation time of an object \citep[due to strong tides,][]{hahn+09,musso+18} and changes its shape, profile and velocity dispersion structure \citep[due to strong external flows,][]{bprg17,mk19}.
Consistently with this picture, tidal influences on substructure also start well before accretion onto the parent object \citep{behroozi+14}.
Similarly, the presence/absence of neighbours having larger \citep{hahn+09,hbv16,salcedo+18} or comparable mass \citep{johnson+19}, and their corresponding tidal influence, has also been shown to be connected with assembly bias.  \citep[See also][for the related effect of tidal heating due to the formation of cosmic sheets.]{myvk05,bh19}

The fact that $\alpha$ simultaneously explains multiple assembly bias trends over a wide range of halo mass suggests that, ultimately, the quantity relevant for assembly bias is the degree of anisotropy of the \emph{current} tidal environment of distinct haloes, evaluated at the current turn-around scale ($\sim4\,\times\,$the halo radius). Having fixed this, the specific physical mechanism that affects any particular variable becomes less relevant; we expect it to only play a role in establishing how strongly that variable correlates with the tidal anisotropy.

This has consequences of practical interest, particularly because $\alpha$ is defined at intermediate length scales. On the one hand, the importance of $\alpha$ as an assembly bias indicator might be exploited to populate low-resolution simulations with otherwise unresolved haloes having the correct assembly bias trends. This would be of immense interest for precision cosmological analyses that would otherwise require high dynamic range as well as tight control on assembly bias related systematics \citep[see, e.g.,][]{zhv14}. On the other hand, $\alpha$ can also be useful in \emph{high} resolution, small volume simulations of galaxy formation, where it might be used to predict (albeit with large scatter) the large-scale environment of realistic galaxies. For example, understanding the strength and origin of correlations between $\alpha$ and variables such as stellar mass, star formation rate, metallicity, etc., might help in understanding the expected strength of \emph{galaxy} assembly bias, which has been difficult to detect robustly in observational samples \citep{lin+16,tinker+17}.

To try and understand \emph{why} the variable $\alpha$, specifically, is such a good assembly bias indicator for distinct haloes, it is worth considering its behaviour for \emph{splashback} haloes. As we showed, $\alpha$ \emph{does not perform well} in explaining the assembly bias of these objects. This is likely a manifestation of the fact that the internal properties of splashback objects, like other substructure, have been dramatically affected by the strong tidal influence of their \emph{host} halo. Since this also includes substantial mass loss due to tidal stripping and a consequent decrease in radius, it is perhaps not surprising that the tidal environment evaluated at the scale $\sim4\,\times\,$the current radius, at the current location, is not a good indicator of the large-scale environment of the splashback object.

It appears, then, that $\alpha$ is a good indicator of assembly bias for objects whose current tidal environment is the most extreme they have ever experienced, and fails for objects whose current environment does not reflect the largest tidal influences that have acted on them. This points towards a novel approach in thinking about substructure in general, in which haloes might be classified by their \emph{tidal history}. Objects that have always been in tidally mild, isotropic environments (small $\alpha$) would then be distinguished from objects that have spent a considerable fraction of their existence in anisotropic sheets or filaments (large $\alpha$). Subhaloes and splashback objects would then simply be the extremes of the latter category, objects that have experienced very high tidal forces at some point in their past (not necessarily reflected by their current environment). Of course, for this picture to be consistent, it must also be possible to construct a local tidal indicator of large-scale assembly bias trends for subhaloes and splashback objects, perhaps $\alpha$ defined using the scale of the \emph{host} halo.

We also believe these ideas could be a useful starting point for a dynamical model of the influence of local non-linear tides on internal halo properties, building on, e.g., known results from tidal torque theory for the connection between large-scale tides and halo angular momenta and shapes \citep[see, e.g.][]{ct96a} and accounting for known correlations between internal halo properties \citep[see, e.g.,][]{sm11,jeeson-daniel+11}.
Finally, it would be interesting to extend our analysis to include tensor assembly bias signatures involving alignments between the mass/velocity ellipsoid tensors, angular momenta and halo-centric tidal tensors. We will return to all these issues in future work.
 
\section*{Acknowledgments}
We thank R. Srianand, K. Subramanian, T. Padmanabhan and S. More for useful discussions, and the anonymous referee for a detailed and helpful report.
The research of AP is supported by the Associateship Scheme of ICTP, Trieste and the Ramanujan Fellowship awarded by the Department of Science and Technology, Government of India. 
OH acknowledges funding from the European Research Council (ERC) under the European Union’s Horizon 2020 research and innovation programme (Grant Agreement No. 679145, project ``COSMO-SIMS'').
We gratefully acknowledge the use of high performance computing facilities at IUCAA, Pune.

 \bibliography{masterRef}

\appendix

\section{Convergence and additional tests}
\label{app:alphaVsdelta}
\noindent
In this Appendix, we first present a convergence study for our calculation of tidal variables which justifies our choices for the minimum halo mass threshold in our simulations. We then show that, although the variables $\alpha$ and $\delta$ defined in the main text are correlated, the tidal anisotropy $\alpha$ is likely to be a better indicator than the isotropic overdensity $\delta$ of all assembly bias, an expectation which is then confirmed in the main text. We also display the 1-dimensional probability distributions of all the halo-related variables used in this work, in a few narrow mass ranges. Finally we complete our analysis in section~\ref{subsec:robustness} by showing explicitly the structure of distribution of {$b_1$,$\alpha$ and $c$} for all $c\in\{\beta,c_v/a_v,c/a,c_{\rm vir},\lambda\}$.

\begin{figure}
\centering
\includegraphics[width=0.45\textwidth]{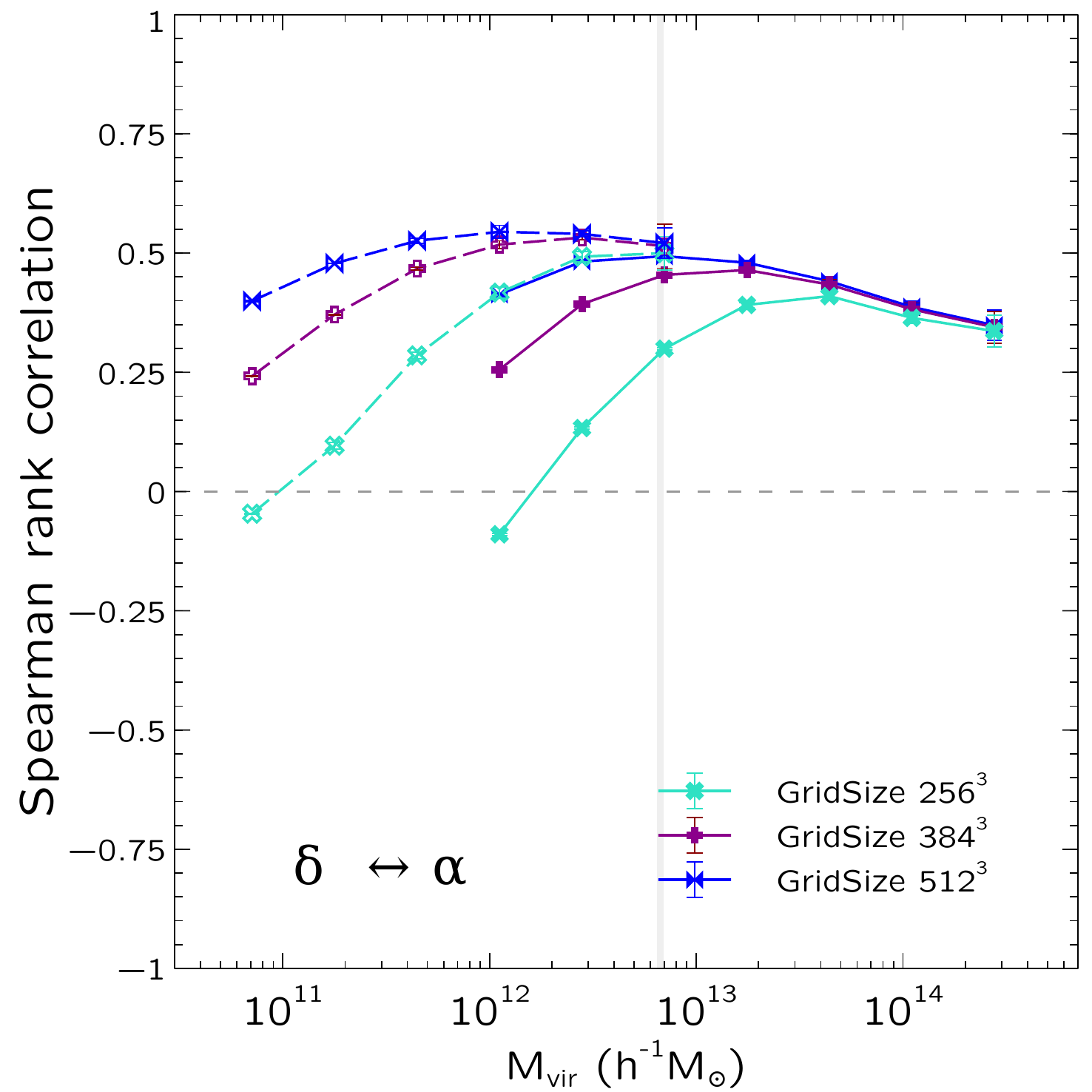}
\caption{Convergence study of the $\alpha\leftrightarrow\delta$ correlation. The symbols joined by lines of different colours indicate measurements using $\alpha$ and $\delta$ (section~\ref{subsec:tidalmsrmnt}) computed on cubic grids of different sizes as indicated.
The formatting of symbols (filled versus empty) and lines (solid versus dashed) is identical to that in Figure~\ref{fig:assemblybias}. Based on the behaviour of the curves in the overlap region between the higher and lower resolution boxes, we conclude that a $512^3$ grid is sufficient for our purposes, provided we restrict attention to haloes with $\geq3200$ particles (shown as the vertical line). See text for further details and a discussion of the consequences of a positive correlation between $\alpha$ and $\delta$.
}
\label{fig:convergence}
\end{figure}

\subsection{Convergence study}
\label{app:convergence}
\noindent
Figure~\ref{fig:convergence} shows the Spearman rank correlation coefficient $\gamma_{\alpha\delta}$ between $\alpha$ and $\delta$ as a function of halo mass. These variables were evaluated as described in section~\ref{subsec:tidalmsrmnt} using various grid sizes as indicated in the legend. Results are shown for the low-resolution (markers with solid lines) and high-resolution configuration (markers with dashed lines). 

\begin{figure}
\centering
\includegraphics[width=0.45\textwidth]{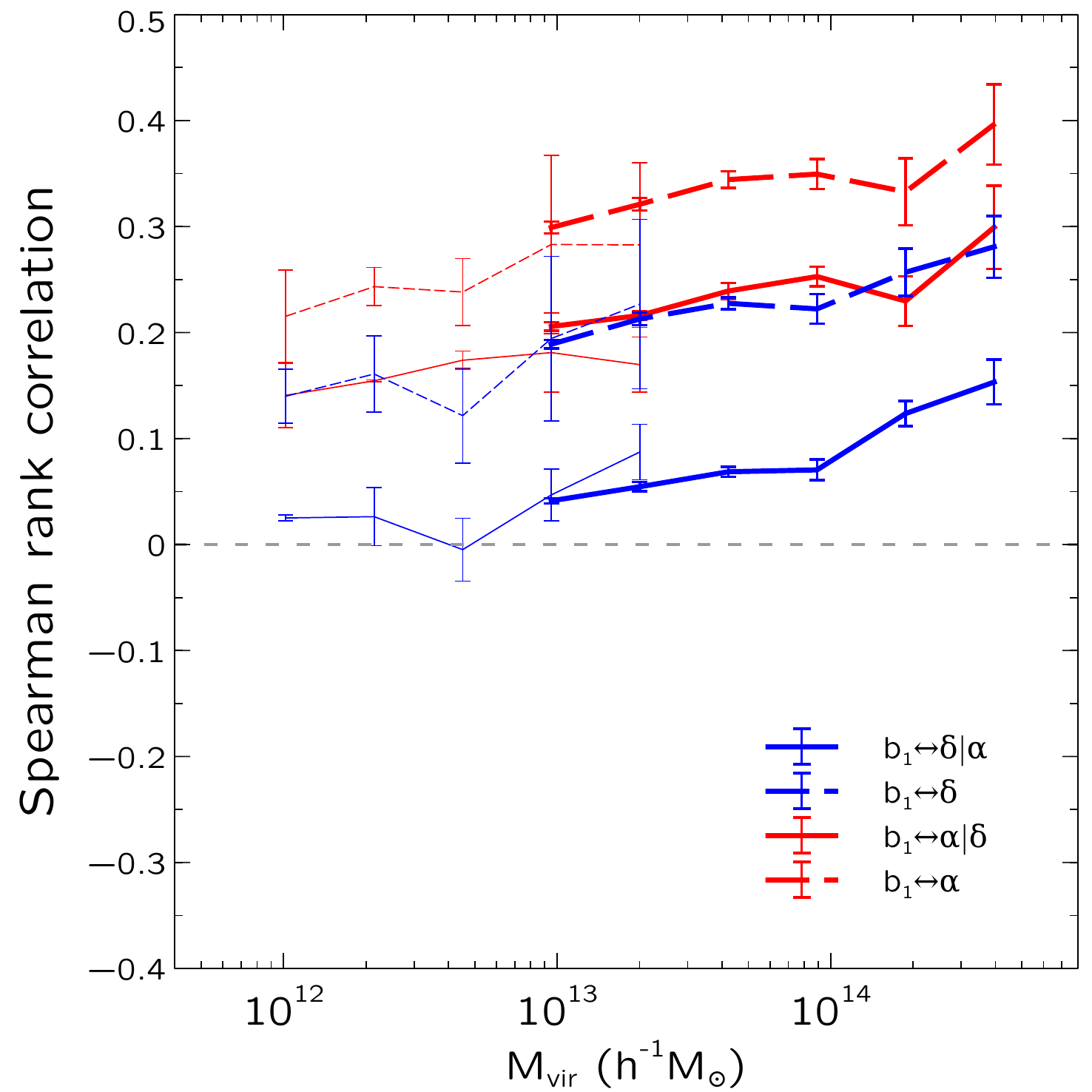}
\caption{Correlation between tidal environment at $4R_{\rm 200b}$ scales (as measured by $\alpha$ and $\delta$) and large-scale environment (measured by halo bias $b_1$). Curves show the unconditional correlation coefficients $\gamma_{b_1\alpha}$ and $\gamma_{b_1\delta}$ (dashed), as well as the conditional coefficients $\gamma_{b_1\delta|\alpha}$ and $\gamma_{b_1\alpha|\delta}$ (solid). The results indicate that $\alpha$ is a better indicator of large-scale environment than is $\delta$, both in the unconditional and conditional sense.}
\label{fig:tidal-bias}
\end{figure}

We see that convergence in any given configuration of the simulations is starting to be achieved at grid sizes of $\geq512^3$ cells. Based on the trends seen in Figure~\ref{fig:convergence}, we also choose a minimum halo mass threshold of $3200$ particles as a compromise between minimising the mismatch in $\gamma_{\alpha\delta}$ between the two configurations and retaining enough statistics in the highest mass bin analysed in the high-resolution simulation. A lower mass threshold would increase the mismatch, while a higher threshold such as $4000$ particles would minimise the mismatch but make all measurements at the high mass end of the high-resolution box too noisy to be reliable.

Since the correlation coefficient between $\alpha$ and $\delta$ is quite large across all masses, one would worry that any statements about statistical connections between $\alpha$ and other variables such as halo bias or internal halo properties could simply be reflecting a correlation between $\delta$ and these properties. Below we demonstrate that this is not the case for any of the correlations we are interested in.

\subsection{Tidal environment and large-scale bias}
\label{app:tidal-bias}
\noindent
Figure~\ref{fig:tidal-bias} explores the correlations between the environment variables $\alpha$ and $\delta$ defined at $\sim4R_{\rm 200b}$ scales and the large-scale environment as measured by halo bias $b_1$. The dashed curves show the unconditional correlation coefficients $\gamma_{b_1\alpha}$ (red) and $\gamma_{b_1\delta}$ (blue). As already discussed by \citet{phs18a}, these show that $\gamma_{b_1\alpha} > \gamma_{b_1\delta}$, so that $\alpha$ is better correlated with $b_1$ than is $\delta$ at any halo mass. Indeed, \citet{phs18a} motivated the choice of $4R_{\rm 200b}$ as being the largest scale (adapted to the halo size) where this is true across all halo masses (see their Figure~5 and also the discussion below).

\begin{figure*}
\centering
\includegraphics[width=0.9\textwidth]{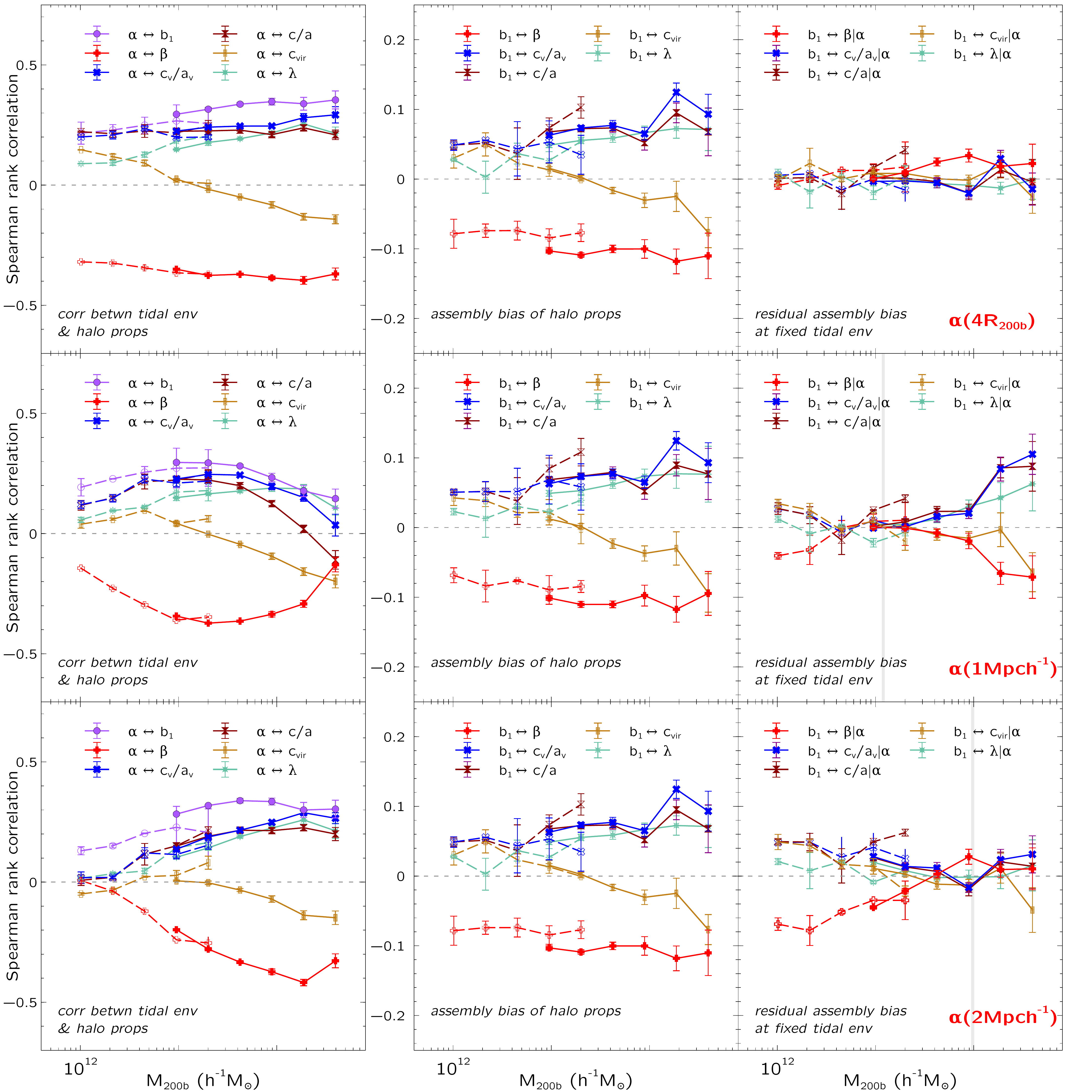}
\caption{The \emph{top panel} is identical to Figure~\ref{fig:corr-main} except that we show results as a function of $M_{\rm 200b}$ instead of $M_{\rm vir}$, with tidal anisotropy $\alpha$ still defined at $4R_{200b}$ scales. The \emph{middle} and \emph{lower panels} repeat the same analysis for $\alpha$ defined at $1\Mpch$ and $2\Mpch$, respectively. We see that the residual assembly bias \emph{(right panels)} is consistent with being zero only when $\alpha$ is defined at scale $4R_{200b}$. In fact, we see that, when $\alpha$ is defined at 1 or 2 \Mpch\ scales, the residual assembly bias reduces only in the mass range where the haloes have $4R_{200b} \simeq$ 1 or 2 \Mpch, respectively. The respective mass ranges have been marked with gray vertical lines in the middle and bottom right panels, respectively (this comparison is the reason to use $M_{\rm 200b}$).
}
\label{fig:corr-main-12R200b}
\end{figure*}

\begin{figure}
\centering
\includegraphics[width=0.45\textwidth]{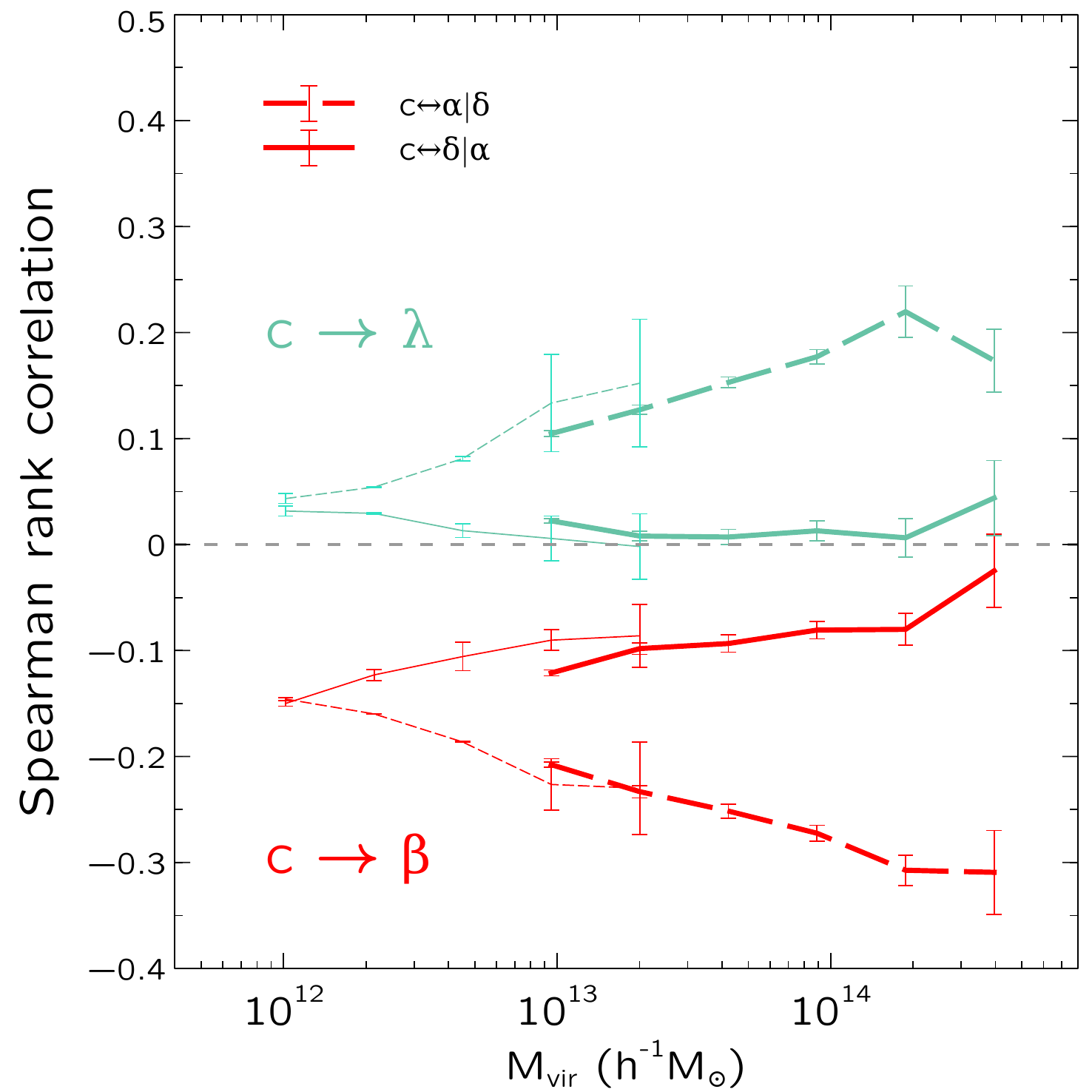}
\caption{Correlation between tidal environment at $4R_{\rm 200b}$ scales (as measured by $\alpha$ and $\delta$) and internal halo properties $c\in\{\beta,\lambda\}$. Curves show the conditional coefficients $\gamma_{c\delta|\alpha}$ (solid) and $\gamma_{c\alpha|\delta}$ (dashed). We see that $|\gamma_{c\delta|\alpha}| < |\gamma_{c\alpha|\delta}|$ at essentially all masses in both cases, indicating that $\alpha$ accounts for a substantial fraction of the correlation of $\delta$ with both of these internal properties. We find qualitatively similar results for the other internal properties $\{c_v/a_v,c/a,c_{\rm vir}\}$ (not shown).}
\label{fig:tidal-internal}
\end{figure}

The solid curves show the conditional correlation coefficients $\gamma_{b_1\alpha|\delta}$ (red) and $\gamma_{b_1\delta|\alpha}$ (blue). We see that $\gamma_{b_1\delta|\alpha} < \gamma_{b_1\delta}$ by a factor $\sim2$-$3$ for all halo masses. The conditional coefficient $\gamma_{b_1\alpha|\delta}$, on the other hand, shows a smaller decrement compared to the correspoding unconditional coeffcient $\gamma_{b_1\alpha}$. In fact, we curiously also see $\gamma_{b_1\alpha|\delta}\simeq\gamma_{b_1\delta}$ across all masses, so that conditioning on $\delta$ does not even decrease the correlation between $\alpha$ and $b_1$ below the \emph{unconditional} correlation between $\delta$ and $b_1$.

\begin{figure*}
\centering
\includegraphics[width=0.8\textwidth]{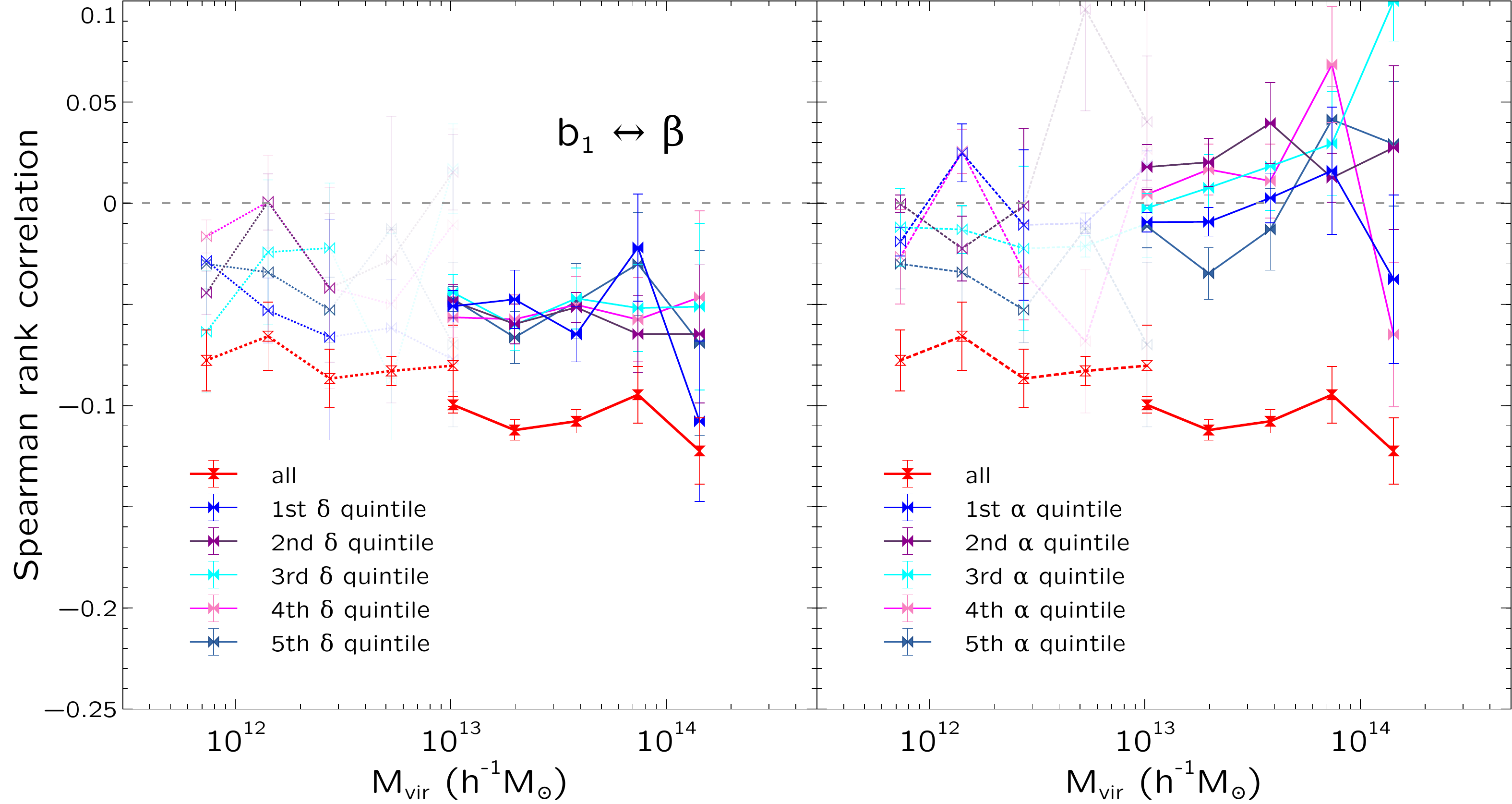}
\caption{Explicit conditional correlation between halo bias $b_1$ and velocity anisotropy $\beta$ as a function of halo mass for haloes in quintiles of $\delta$ \emph{(left panel)} and $\alpha$ \emph{(right panel)}. The all-halo coefficient is shown with red symbols joined by red lines; this is the same in each panel and is repeated from the middle panel of Figure~\ref{fig:corr-main}. We see that fixing $\alpha$ leads to conditional correlations that are substantially closer to zero than when fixing $\delta$.}
\label{fig:corr-betab1-deltaalpha}
\end{figure*}

These results indicate that $\alpha$ is a better indicator of large-scale environment than is $\delta$, both in the unconditional and conditional sense.

We have also repeated the analysis of Figure~\ref{fig:corr-main} using $\alpha$ defined at fixed scales of $1\Mpch$ and $2\Mpch$, finding that the cancellations leading to small conditional correlation coefficients only occur in the mass range where $4R_{\rm 200b}\simeq 1\Mpch,2\Mpch$, respectively. Figure~\ref{fig:corr-main-12R200b} shows the results. For ease of comparison, we use bins of $M_{\rm 200b}$ rather than $M_{\rm vir}$ for this Figure. Finally, we have repeated this last analysis using $\delta$ instead of $\alpha$, defined at $4R_{\rm 200b},1\Mpch,2\Mpch$. We found that none of these variables perform as well as $\alpha(4R_{\rm 200b})$ in producing small conditional correlation coefficients across the entire halo mass range we probe. For brevity, we do not display these results. In a recent study, \citet{han+19} proposed that $\delta$ defined at $1$-$2\Mpch$ is a strong candidate for explaining assembly bias trends. Our results indicate that $\alpha(4R_{\rm 200b})$ is an even stronger candidate, which can also be understood by the fact that the $\alpha\leftrightarrow b_1$ correlation seen in the left panel Figure~\ref{fig:corr-main} takes values at least comparable to, and usually larger than, the correlation strength between $b_1$ and the fixed-scale $\delta$'s (not shown).

\subsection{Tidal environment and internal halo properties}
\label{app:tidal-internal}
\noindent
Figure~\ref{fig:tidal-internal} explores the correlations between internal halo properties and the environmental variables $\alpha$ and $\delta$, colour-coded by the internal properties as in previous Figures. The solid (dashed) curves show the conditional correlation coefficients $\gamma_{c\delta|\alpha}$ ($\gamma_{c\alpha|\delta}$) for $c\in\{\beta,\lambda\}$. We have chosen these two internal variables as representing the extremes of the trends we discuss here; the other internal variables $\{c_v/a_v,c/a,c_{\rm vir}\}$ show qualitatively identical trends with intermediate strengths. 
In each case, we find $\gamma_{c\delta|\alpha}$ is substantially smaller in magnitude than $\gamma_{c\alpha|\delta}$ at all but the smallest halo masses we explore, indicating that $\alpha$ accounts for a substantial fraction of the correlation of $\delta$ with \emph{all} internal properties. Especially in the case of $\lambda$, we see that $\alpha$ accounts for nearly \emph{all} of the correlation between $\lambda$ and $\delta$.

Taken together, the results shown in Figures~\ref{fig:tidal-bias} and~\ref{fig:tidal-internal} show that $\alpha$ is a much better candidate than $\delta$ for an environmental link that could explain assembly bias in any internal halo property. In other words, the anisotropy of the halo tidal environment is expected to be more important than the local density in explaining assembly bias trends. 

\subsection{Explicit conditional correlation}
\label{app:explicitcondcorr}
\noindent
In the main text, we explore the connection between halo tidal environment and assembly bias using the Gaussian-motivated correlation coefficients $\gamma_{bc|a}\equiv\gamma_{bc}-\gamma_{ab}\gamma_{ac}$, where $b$ and $c$ represent halo bias and any internal halo property, respectively, and $a$ represents the environmental variable. Here, we perform an explicit test of this connection by evaluating correlation coefficients in \emph{fixed bins} of the environmental variable. Since binning naturally increases the noise in our measurements, we only display results for the strongest assembly bias trend which is that between $b_1$ and velocity anisotropy $\beta$.

Figure~\ref{fig:corr-betab1-deltaalpha} shows the correlation coefficients $\gamma_{b_1\beta}$ as a function of halo mass, evaluated for haloes in quintiles of $\delta$ \emph{(left panel)} and $\alpha$ \emph{(right panel)}, with the all-halo coefficient repeated in each panel in red. It is visually apparent that fixing $\alpha$ leads to conditional correlations that are substantially closer to zero than when fixing $\delta$. We have checked that qualitatively similar results hold for all other internal variables except $c_{\rm vir}$ for which the noise is too large to draw strong conclusions given our simulation set.

\subsection{Halo properties}
\label{app:internal}
\noindent
Here, we show for reference the distributions of all variables studied in the main text, including halo bias $b_1$, environmental variables $\{\alpha,\delta\}$, and internal halo properties $\{\beta,c_v/a_v,c/a,c_{\rm vir},\lambda\}$. See section~\ref{sec:sims} for a description of how each of these is measured.

\begin{figure*}
\centering
\includegraphics[width=\textwidth]{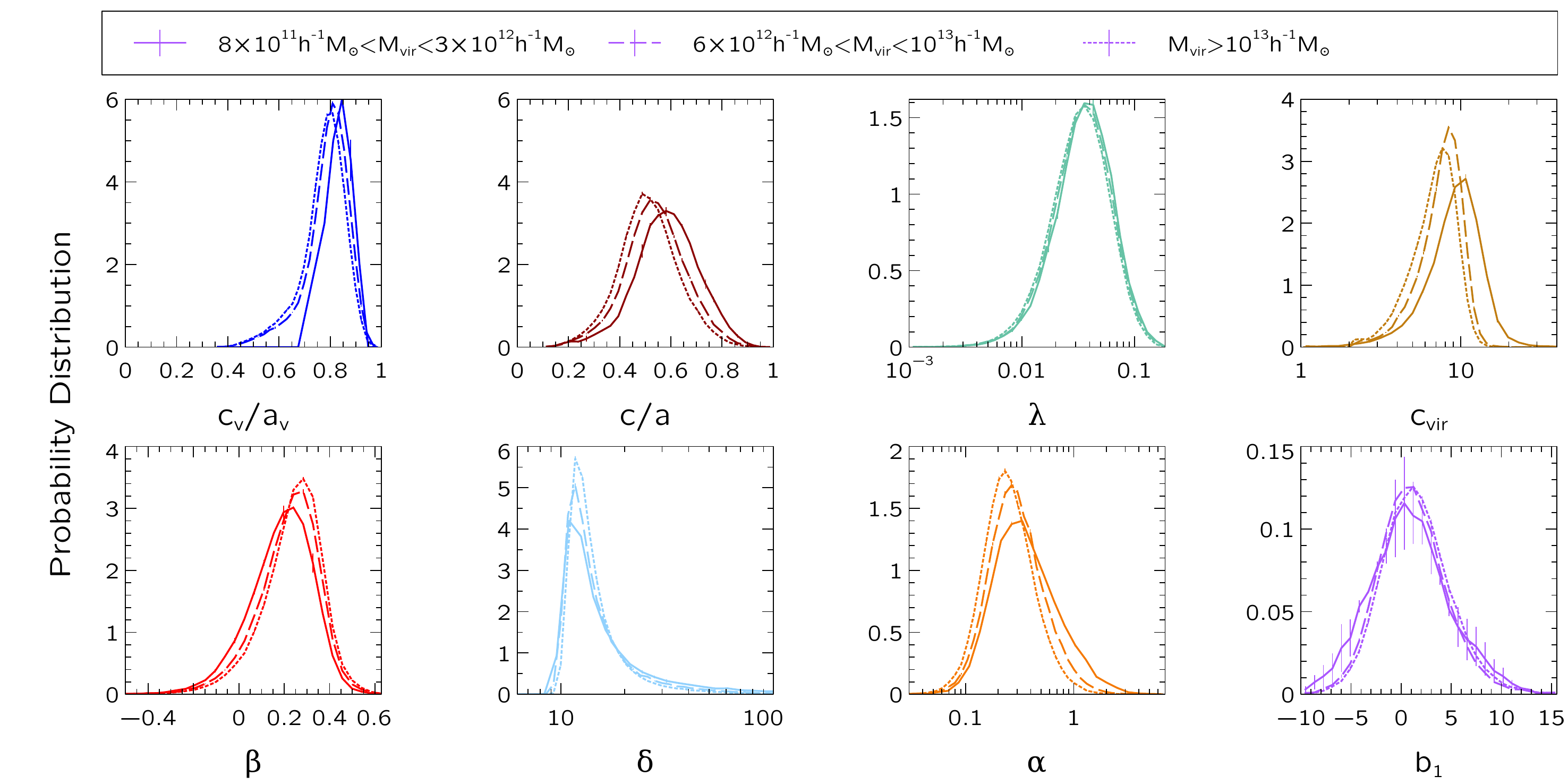}
\caption{Normalised distributions of all variables used in this analysis (different panels, as indicated). In each panel, the solid, dashed and dotted curves show measurements for low, intermediate and high mass haloes, respectively, as indicated in the legend at the top. The lowest mass bin uses measurements from 2 realisations of the high resolution box, while the other two mass bins use measurements from 10 realisations of the low resolution box, with the curves showing the mean and the error bars showing the scatter across realisations.}
\label{fig:histogram}
\end{figure*}

The histograms in Figure~\ref{fig:histogram} show the individual distributions of all $8$ variables (different panels, as labelled) for a few narrow mass ranges (different line styles), with several known trends being apparent. 
We see that haloes are, on average, substantially aspherical in shape (panel $c/a$) but less so in their velocity ellipsoids (panel $c_v/a_v$), although there is a clear preference for radially dominated orbits (panel $\beta$). The distributions of spin $\lambda$ and concentration $c_{\rm vir}$ show distinct tails at small values, while those of the environmental variables $\delta$ and $\alpha$ are skewed towards large values, and the distributions of $b_1$ are largely symmetric around the median. All variables except $b_1$ and $\lambda$ show noticeable trends with halo mass.

\subsection{Joint distribution of $\alpha$, $b_1$ and internal properties}
\label{app:jointdistribution}
\noindent
In section~\ref{subsec:robustness} we analysed the full distribution of {$b_1$, $\alpha$, $\beta$} and showed that
the overall anti-correlation between $b_1$ and $\beta$ is consistent with being largely due to $\alpha$. In this section we complete this analysis by showing the same 
for the distribution of $b_1$, $\alpha$ and $c$ for all internal properties $c$ in all three mass ranges.

\begin{figure*}
\centering
\includegraphics[width=\textwidth]{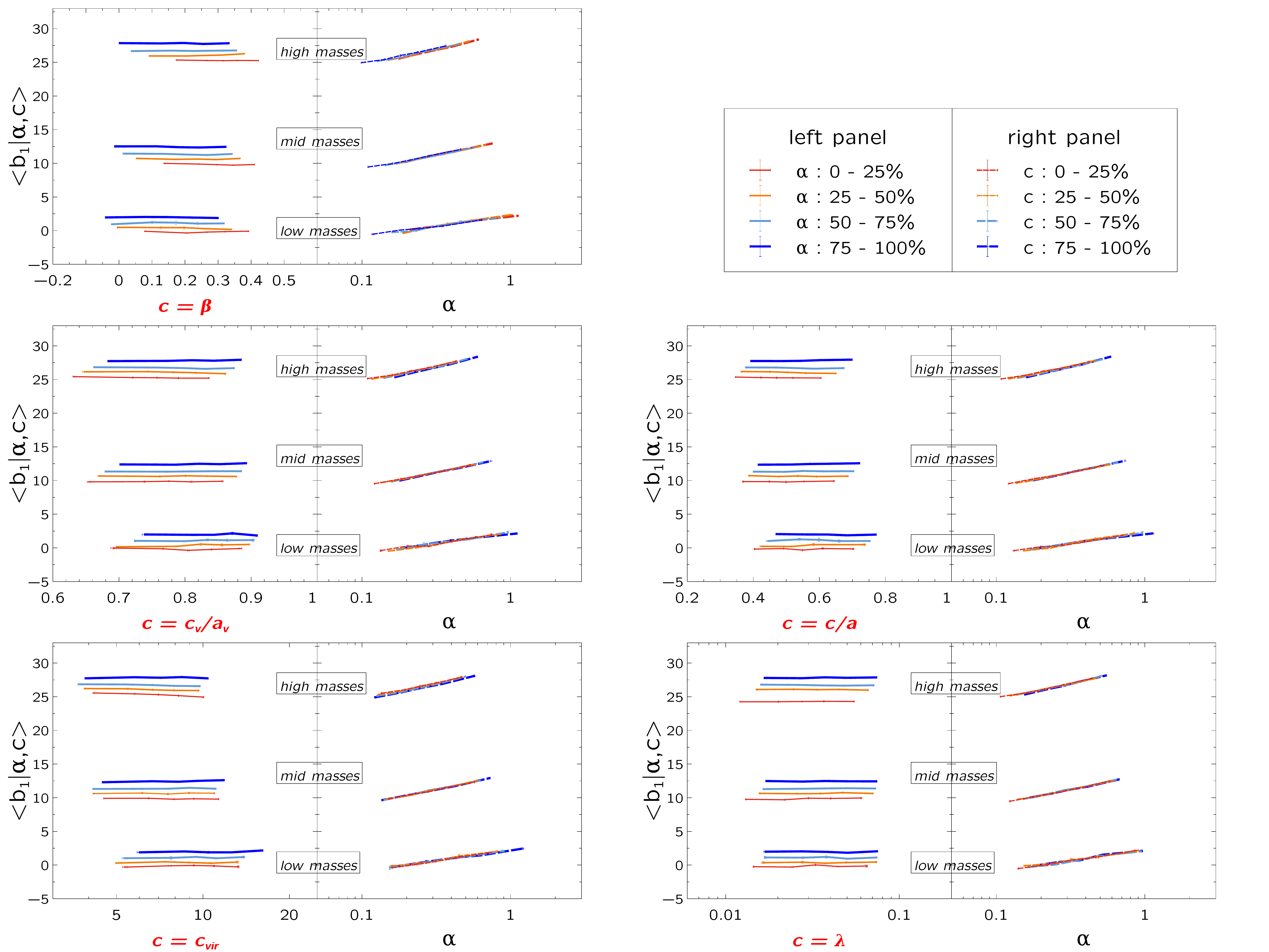}
\caption{{\bf Conditional mean of $b_1$ at fixed $\alpha$ and $c\in\{\beta,c_v/a_v,c/a,c_{\rm vir},\lambda\}$:} Each subplot shows $\avg{b_1|\alpha,c}$ in two projections: the $b_1$-$c$ plane for diffrent quartiles of $\alpha$ (coloured lines in the \emph{left subplot panels}) and the $b_1$-$\alpha$ plane for different quartiles of $c$ (coloured lines in the \emph{right subplot panels}). The mapping between line colour and quartiles of $\alpha$ or $c$ is given in the legend at the top right.
The results for $c=\beta$ are repeated from Figure~\ref{fig:dist-betab1alpha}. 
In each panel, results are shown separately for haloes segregated into three mass ranges as in Figure~\ref{fig:dist-betab1alpha}. 
The low-mass results are averaged over $2$ realisations of the high resolution box while the mid- and high-mass results are averaged over $10$ realisations of the low resolution box, with error bars in each case showing the error on the respective mean.
For clarity, the mid- and high-mass results are also given vertical offsets of $+10$ and $+25$, respectively.  
We see that the results for all internal variables $c$ and for each mass range are consistent with the relation $\avg{b_1|\alpha,c}\simeq\avg{b_1|\alpha}$. 
See text for a discussion.}
\label{fig:bias_mean}
\end{figure*}

\begin{figure*}
\centering
\includegraphics[width=0.98\textwidth]{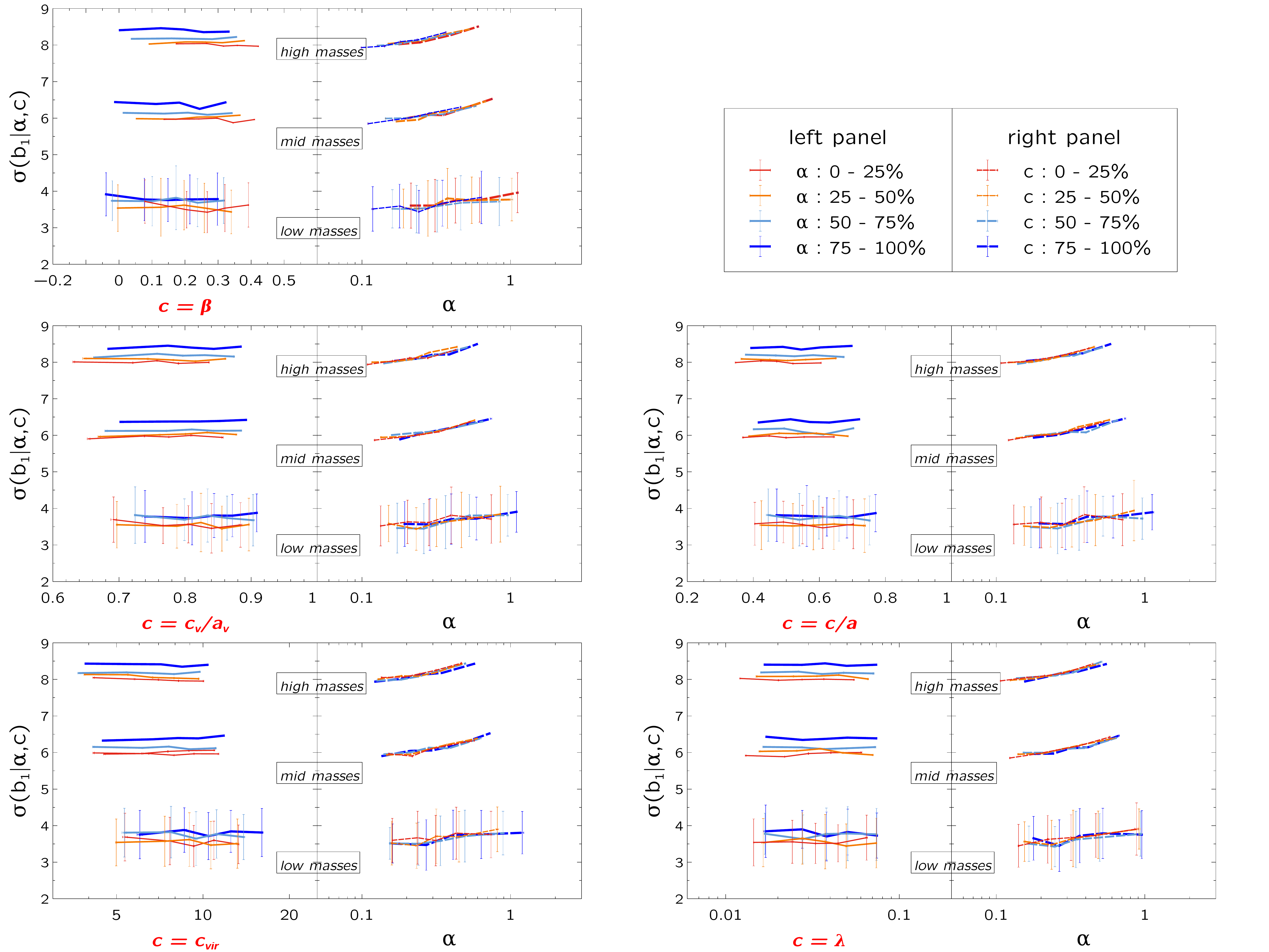}
\caption{{\bf Square-root of conditional variance of $b_1$ at fixed $\alpha$ and $c\in\{\beta,c_v/a_v,c/a,c_{\rm vir},\lambda\}$}: Same as Figure~\ref{fig:bias_mean}, showing results for $\sigma(b_1|\alpha,c) \equiv (\avg{b_1^2|\alpha,c}-\avg{b_1|\alpha,c}^2)^{1/2}$.
In this case, the mid- and high-mass results in each subplot panel were given vertical offsets of $+3$ and $+5$, respectively, for clarity.
The large errors in the low-mass results are likely driven by systematic effects in computing object-by-object $b_1$ values due to the smaller $k$-space range provided by the smaller volume of the high resolution box (see also the low-mass histogram of $b_1$ in Figure~\ref{fig:histogram}).
We see that the results for all internal variables $c$ and for each mass range are consistent with the relation $\sigma(b_1|\alpha,c)\simeq\sigma(b_1|\alpha)$.
Together with the results of Figure~\ref{fig:bias_mean}, this shows that the full distribution of $b_1$, $\alpha$ and $c$ is consistent with $p(\alpha,b_1,c)\simeq p(\alpha)p(b_1|\alpha)p(c|\alpha)$ for each $c$. See text for a discussion. 
}
\label{fig:bias_variance} 
\end{figure*}

To make the results compact, we will not show scatter plots for the various distributions and instead focus on the conditional mean $\avg{b_1|\alpha,c}$ (as already shown in Figure~\ref{fig:dist-betab1alpha} for the case $c\to\beta$) and additionally the square-root of the conditional variance $\sigma(b_1|\alpha,c) \equiv \left(\avg{b_1^2|\alpha,c}-\avg{b_1|\alpha,c}^2\right)^{1/2}$, for all $c\in\{\beta,c_v/a_v,c/a,c_{\rm vir},\lambda\}$. If the general relation \eqref{eq:mainresult} is true, then we should expect $\avg{b_1|\alpha,c}\simeq\avg{b_1|\alpha}$ and $\sigma(b_1|\alpha,c)\simeq\sigma(b_1|\alpha)$. Figures~\ref{fig:bias_mean} and~\ref{fig:bias_variance} show that this is indeed the case, as we discuss below.

Both Figures comprise of subplots focusing on one internal variable $c$ at a time. Figure~\ref{fig:bias_mean} (Figure~\ref{fig:bias_variance}) shows projections of $\avg{b_1|\alpha,c}$  ($\sigma(b_1|\alpha,c)$) in the $b_1$-$c$ (left subplot panels) and $b_1$-$\alpha$ planes (right subplot panels) for all three mass ranges (three sets of curves with offsets given for clarity). 
The bins along each horizontal axis are chosen as quintiles of the respective variable,\footnote{The marker location on the horizontal axis for each such quintile is chosen as the median of that quintile, averaged over realisations.} so that the left subplot panels additionally reveal the $b_1$-$c$ assembly bias trends for the mean and width of the conditional $b_1$ distributions as systematic horizontal shifts of the different lines.
Figure~\ref{fig:bias_mean} for $\avg{b_1|\alpha,c}$ shows results qualitatively identical to those seen in Figure~\ref{fig:dist-betab1alpha}, with the $b_1$-$c$ projections being approximately horizontal lines in fixed quartiles of $\alpha$, and the $b_1$-$\alpha$ projections tracing out common loci in fixed quartiles of $c$. Figure~\ref{fig:bias_variance} extends these results to the (square-root of) conditional variance  $\sigma(b_1|\alpha,c)$, with the $b_1$-$c$ projections again being approximately horizontal with $\alpha$-dependent offsets, and the $b_1$-$\alpha$ projections tracing out common loci in fixed $c$-quartiles.

Furthermore, as we see in Figure~\ref{fig:histogram}, the distribution of $b_1$ is approximately symmetric about its mean, indicating that a Gaussian shape is a reasonable approximation. This would mean that the conditional independence of $b_1$ and $c$ at fixed $\alpha$ as seen in the conditional mean and variances in fact extends to the entire distribution as discussed above. These results strongly support our main conclusions regarding the role of $\alpha$ in explaining assembly bias.

\label{lastpage}

\end{document}